\documentclass[12pt]{article}
\usepackage{graphicx}
\usepackage{epsfig}
\usepackage{latexsym}
\usepackage{latexsym}
\usepackage{amssymb}
\newcommand{\labell}[1]{\label{#1}}
\newcommand{\reef}[1]{(\ref{#1})}

\newcommand{\bibbyitem}[1]{\bibitem{#1}}

\def\f{{\hat f}}

\def\etc{{\it etc}}

\def\eg{{\it e.g.}}

\def\ie{{\it i.e.}}

\DeclareSymbolFont{AMSb}{U}{msb}{m}{n}
\DeclareMathSymbol{\IN}{\mathbin}{AMSb}{"4E}
\DeclareMathSymbol{\IZ}{\mathbin}{AMSb}{"5A}
\DeclareMathSymbol{\IR}{\mathbin}{AMSb}{"52}
\DeclareMathSymbol{\Q}{\mathbin}{AMSb}{"51}
\DeclareMathSymbol{\II}{\mathbin}{AMSb}{"49}
\DeclareMathSymbol{\IC}{\mathbin}{AMSb}{"43}
\DeclareMathSymbol{\IP}{\mathbin}{AMSb}{"50}
\DeclareMathSymbol{\IH}{\mathbin}{AMSb}{"48}
\DeclareMathSymbol\IA{\mathalpha}{AMSb}{"41}
\DeclareMathSymbol\IS{\mathalpha}{AMSb}{"53}

\def\Q{{\cal Q}}

\begin{document}


\hbox{$\phantom{.}$}

\bigskip
\bigskip
\begin{center}
   {\Large \bf  Oblate, Toroidal, and Other Shapes for the Enhan\c con}


   \end{center}

\bigskip
\bigskip
\bigskip
\bigskip

\centerline{\bf Lisa M. Dyson$^a$, Laur J\"arv$^{b,}$\footnote{Also:
    Institute of Theoretical Physics, University of Tartu, Estonia},
  Clifford V. Johnson$^c$}

\bigskip
\bigskip
\bigskip

  \centerline{\it ${}^{a}$Center
for Theoretical Physics}
  \centerline{\it Department of Physics}
\centerline{\it Massachusetts Institute of Technology}
\centerline{\it Cambridge, MA 02139, U.S.A}

\bigskip
\bigskip

\centerline{\it ${}^{b,c}$Centre
for Particle Theory}
  \centerline{\it Department of Mathematical Sciences}
\centerline{\it University of
Durham}
\centerline{\it Durham, DH1 3LE, U.K.}

\centerline{$\phantom{and}$}

\bigskip

\centerline{\small \tt
 ldyson@ctp.mit.edu,
  laur.jarv@durham.ac.uk,
 c.v.johnson@durham.ac.uk}

\bigskip
\bigskip
\bigskip


\begin{abstract}
  We present some results of studying certain axially symmetric
  supergravity geometries corresponding to a distribution of BPS
  D6--branes wrapped on K3, obtained as extremal limits of a rotating
  solution. The geometry's unphysical regions resulting from the
  wrapping can be repaired by the enhan\c con mechanism, with the
  result that there are two nested enhan\c con shells.  For a range of
  parameters, the two shells merge into a single toroidal surface.
  Given the quite intricate nature of the geometry, it is an
  interesting system in which to test previous techniques that have
  been brought to bear in spherically symmetric situations.  We are
  able to check the consistency of the construction using supergravity
  surgery techniques, and probe brane results.  Implications for the
  Coulomb branch of (2+1)--dimensional pure $SU(N)$ gauge theory are
  extracted from the geometry. Related results for wrapped D4-- and
  D5--brane distributions are also discussed.

\end{abstract}
\newpage \baselineskip=18pt \setcounter{footnote}{0}

\section{Introductory Remarks}

When studying gauge/gravity dualities, one may encounter singularities
in the supergravity geometry.  Some of these singularities are
acceptable, in the sense of having a physical understanding such as
the location of a source (\eg, a brane) in the geometry. Others are
unphysical, and signal a failure of supergravity to capture crucial
features of the situation, such as physics of the underlying
short--distance theory.

One mechanism for resolving such singularities that has appeared in
this context is the ``enhan\c con'' mechanism, so called because the
prototype example\cite{jpp} was accompanied by the appearance of extra
massless states giving rise to enhanced gauge symmetry in spacetime.
The entire supergravity solution came from wrapping many BPS
D6--branes on K3, and the dangerous ``repulson''\cite{repulson}
singularities are associated to a region of repulsive geometry, whose
behaviour is inconsistent with the 1/4--BPS nature of the
configuration.  The resolution of the geometry's singularities was to
simply excise the region which was behaving poorly and replace it with
flat space.

\begin{figure}[ht]
  \centering
  \scalebox{0.31}{\includegraphics[angle=0]{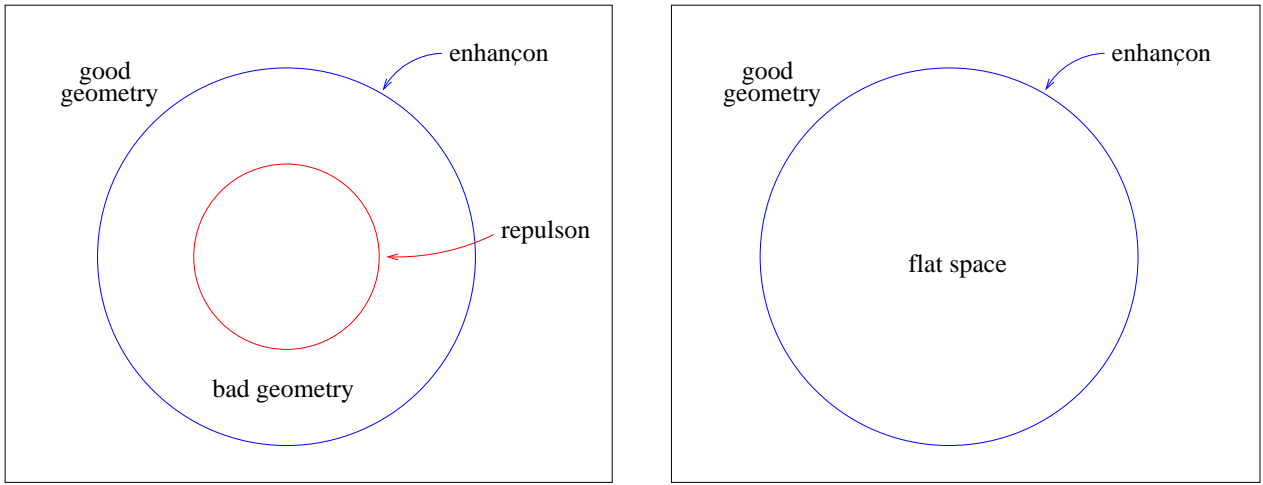}}
\caption{\footnotesize 
  Left: A slice through the enhan\c con (exterior) and repulson
  (interior) loci. Right: The geometry after excision. (The situation
  is spherically symmetric and the three dimensional loci are
  constructed by revolving the plot about the vertical axis.)}
\label{sphere}
\end{figure}

The physics behind this is the fact that all of the branes making up
the solution cease to be pointlike and smear out to form a spherical
shell called the ``enhan\c con'', of radius $r_{\rm e}$, which is
larger than the radius, $r_{\rm r}$ of the repulson singularity. (See
figure~\ref{sphere}.) This is consistent with the fact that the
wrapped branes also play the role of BPS monopoles charged under the
$U(1)$ arising from reduction of the superstring theory on the K3's
volume cycle. In supergravity, at radius~$r_{\rm e}$ the K3's volume
reaches the value at which an effective Higgs vacuum expectation value
 vanishes and the $U(1)$ is restored to $SU(2)$. As BPS
monopoles\footnote{Of course only the BPS one--monopole solution is
  exactly spherical\cite{Bogomolny:1975de}. However, at large~$N$,
  there is no problem finding an approximately spherical
  configuration. The deviation from this spherical symmetry should be
  subleading in a $1/N$ expansion\cite{j1,j}.}, the D6--branes
accordingly smear out, and become massless, forming the enhan\c con
shell. Since there are no longer any point sources inside~$r_{\rm e}$,
the spacetime geometry is well approximated by flat space.

The resulting complete geometry is well--behaved, possessing the
correct physical properties to match certain gauge theory phenomena
expected from the world--volume theory on the branes, such as the
metric on moduli space. The configuration has eight supercharges,
which is enough to have a moduli space, but not so much as to force it
to be trivially flat.

The proposal to excise the bad region and replace it with flat space,
while a natural one, might have seemed considerably drastic from other
points of view, and one might ask whether it is a consistent procedure
from the purely supergravity perspective: Since the enhanced gauge
symmetry is a purely stringy phenomenon, is there any sense in which
the excision procedure is natural in supergravity? After doing a
purely supergravity analysis\cite{jmpr}, the satisfying answer is that
the excision is not only allowed by the supergravity (in fact, one can
perform it at any radius greater than $r_{\rm e}$ in that spherical
case and get sensible results), but it is extremely natural to carry
it out at the enhan\c con radius. The reason is as follows: One glues
the exterior geometry onto the new interior (flat space) and any
mismatch in the extrinsic curvatures of the geometries across the
junction acts as a source in the theory, which in this case would be
the smeared shell of branes.  The analysis of ref.\cite{jmpr} (see
also ref.\cite{maeda}) showed that the shell of branes was in fact
massless at the enhan\c con radius, consistent with the superstring
expectations given above.  Furthermore, the enhan\c con radius was the
{\it most economical} place at which to perform the excision, since
that radius also coincided with the outermost reaches of the
unphysical repulsive interior region, as could be seen by probing with
ordinary supergravity test particles.

While this special spherically symmetric situation is compelling, it
is interesting to find more examples, and explore the nature of the
mechanism in greater detail. The spherical symmetry is quite seductive, and it
is easy to forget that there should be nothing particularly special
about that symmetry. After all, the $N$ constituent branes are BPS,
and so in fact as long as we do not develop any unphysical regions,
{\it very many} shapes should be possible, and for large $N$, almost
{\it any} shape is imaginable. The difficulty is of course in finding
supergravity techniques which facilitate the exhibition of solutions
describing the non--trivial shapes that we can imagine\footnote{We
  understand that the forthcoming work\cite{rob} will also present a
  discussion of the non--spherical case. See also
  ref.\cite{Wijnholt:2001us} for discussions of attempts to embed
  exact multi--monopole solutions into the problem to find
  non--spherical solutions.}.

\begin{figure}[ht]
\scalebox{0.5}{\includegraphics{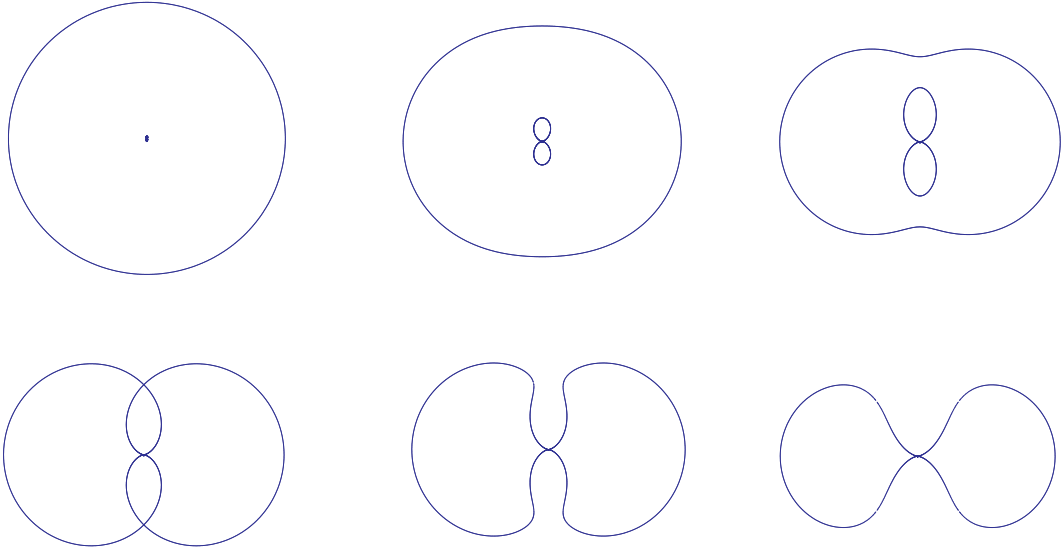}}
 \caption{\footnotesize A series of slices  through the enhan\c con loci for 
   varying values of a parameter $\ell$.  The three dimensional loci
   are constructed by revolving the plots about the vertical axis. The
   first one is closest to the spherical case, $\ell=0$ in
   figure~\ref{sphere}. Here, there is a small interior enhan\c con
   locus at the origin.  For $0<\ell<\ell^{\rm cr}_{\rm e}$, the
   enhan\c con locus is in general disconnected, the interior shape
   being steadily more visible with increasing $\ell$, and the outer
   shell becomes increasingly oblate until at $\ell=\ell^{\rm cr}_{\rm
     e}$, they touch, forming a single shape, which persists as a
   torus for $\ell>\ell_{\rm cr}^{\rm e}$. (Later, we shall see that
   the point in the centre is indeed a hole.)}
\label{loci}
\end{figure}

For this project we set out to do the most simple BPS deviation from
the sphere we could think of, which was to deform to an oblate
situation, and perhaps work perturbatively away from the spherical
situation. To our surprise, the result was much richer than we could
have hoped for. We succeed in describing a complete family of axially
symmetric geometries, parametrised by a parameter~$\ell$. It
transpires that there is a critical value, $\ell^{\rm cr}_{\rm
  e}$, separating two distinct physical situations. The reader should
refer to figure~\ref{loci} (on page~\pageref{loci}) which illustrates
the following descriptions:

\begin{itemize}
\item For $\ell<\ell^{\rm cr}_{\rm e}$, the enhan\c con locus is
  disconnected: There is an outer shell and an inner shell, with
  non--trivial supergravity inside the inner shell. The region between
  the two shells is unphysical in the naive geometry and is excised
  and replaced by flat space, while the region inside the inner shell
  is perfectly well--behaved, and so remains.  For increasing $\ell$
  the outer shell becomes more depressed at the poles, forming an
  oblate shape, while the inner shell reaches increasingly outwards
  towards the poles.
  
\item Even more remarkably, perhaps, for $\ell=\ell^{\rm cr}_{\rm e}$,
  the two enhan\c con shells join, and the enhan\c con is no longer
  disconnected. For $\ell>\ell^{\rm cr}_{\rm e}$, the complete enhan\c
  con locus is in fact a torus. The torus flattens increasingly for
  greater $\ell.$ See also figure~\reef{notsphere}.

\item Another choice which can be made for $\ell<\ell^{\rm cr}_{\rm
    e}$ is to excise the entire interior, leaving the oblate shape
  outside, and flat space in the entire interior region.

\end{itemize}
Remarkably, the supergravity excision technology is tractable in this
situation, and confirms that these are physically sensible choices.

\begin{figure}[ht]
  \centering
\scalebox{0.5}{\includegraphics{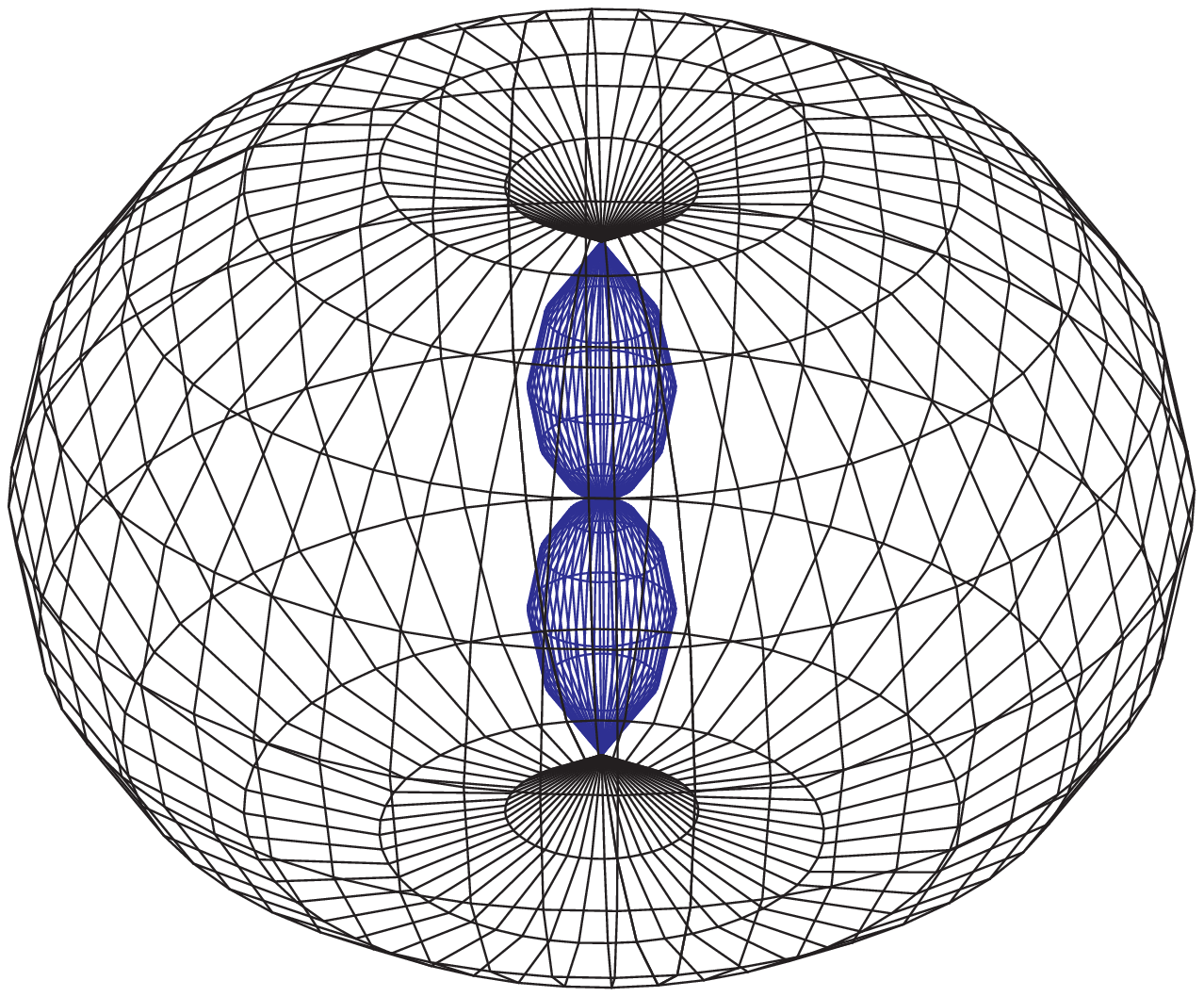}}
\raisebox{0.25cm}{\scalebox{0.44}{\includegraphics{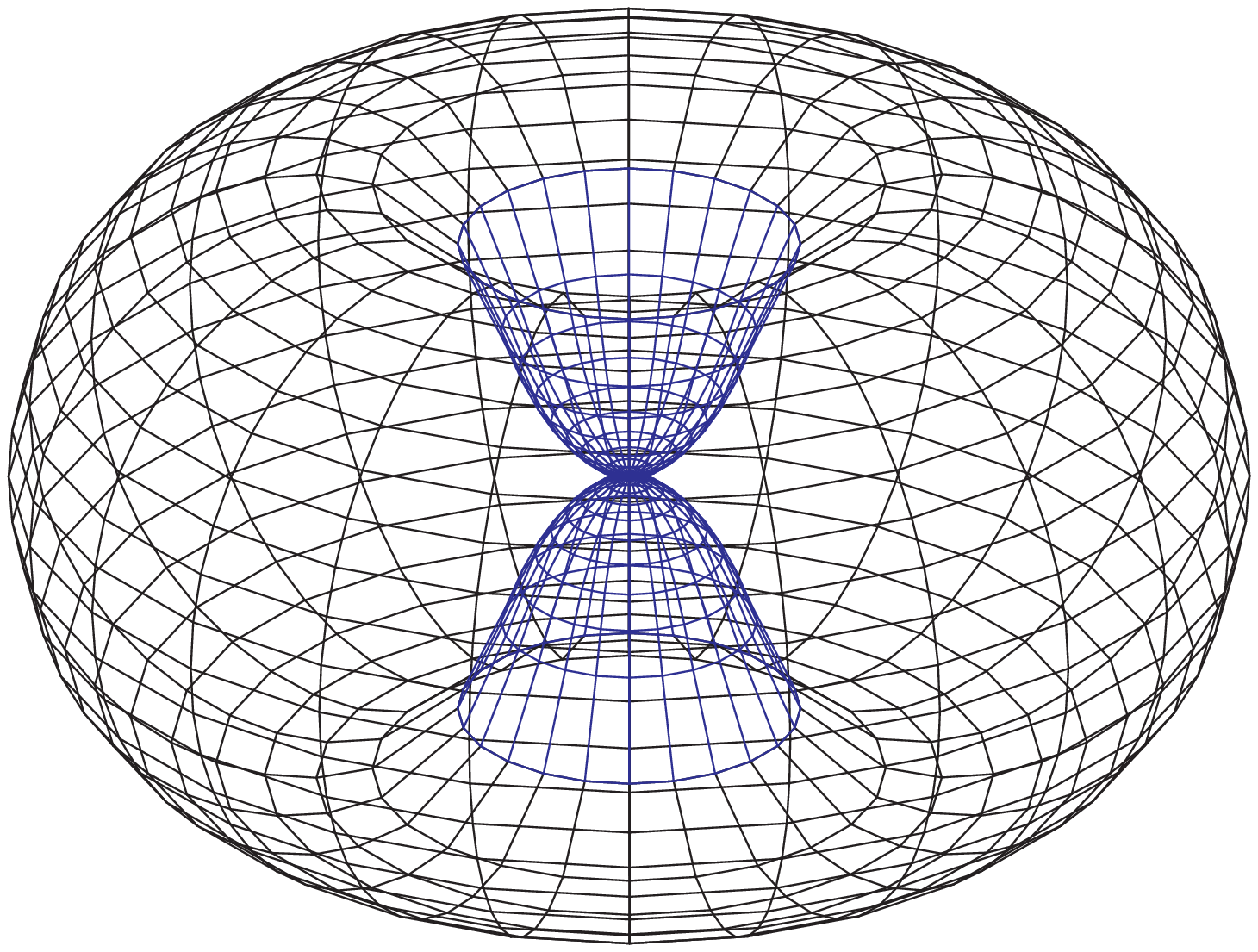}}}
\caption{\footnotesize
  A three dimensional depiction of the connected  enhan\c con locus, at
  $\ell=\ell^{\rm cr}_{\rm e}$ (left), and $\ell>\ell^{\rm cr}_{\rm
    e}$ (right).}
\label{notsphere}
\end{figure}

In this paper we report on this interesting and highly non--trivial
example of supergravity geometry as an exhibit of the enhan\c con
mechanism at work, and as a study which allows us to learn more about
certain brane configurations. In section~\ref{distribution}, we study
first a generalised six--brane supergravity solution which corresponds
to a disc distribution of D6--branes. Such branes are the nicest
geometries to wrap on K3, since the remaining transverse geometry has
three spatial dimensions. This means that the resulting geometry is
easy to visualise, and, at a deeper level, much of the mathematics is
quite familiar, as we shall see. There are two useful coordinate
systems which are employed here to uncover aspects of the geometry in
the wrapped case, and it is worthwhile exploring them in the unwrapped
case. The unwrapped case itself is interesting, and we compare it to
expressions that we derive for other D--brane disc distributions. 

We should make a cautionary note, however. It is useful to describe
these solutions in terms of a continuous density of branes. While this
is a simplification, (since branes come in discrete amounts due to
charge quantisation), it is a good and useful approximation at large
$N$. However, the D6--brane disc distribution that we find has the
strange feature that the density function in the interior of the disc
is in fact negative, which is hard to interpret. However, we are able
to characterise the behaviour resulting from this strange feature and
do not discard the solution, for at least three reasons:
\begin{itemize}
\item Apart from the negative density itself, there is no compelling
  reason to think that there might not be yet to be found some
  beyond--supergravity means of explaining the role of such a
  geometry, perhaps involving something analogous to the enhan\c con
  mechanism\footnote{In fact, one of the choices we can make in the
    resolution does entirely cut out the negative density region, so
    there are some situations in which the enhan\c con can remove the
    negative density, but this cannot be the whole story.}.
  
\item The negative density has {\it nothing } to do with the
  repulson geometry arising from wrapping, which is what we really
  want to study in this paper.  We need not ``throw out the baby with the
  bathwater'', and can carry out a {\it separate} discussion of the
  wrapping and its associated features, including the enhan\c con
  mechanism. The features associated to this negative density are
  clearly isolated in the discussion and do not play any role.
  
\item The D5-- and D4--brane distributions which we also display have
  no such strange behaviour, and the subsequent discussion of the
  appearance of the repulson after wrapping and its resolution by the
  enhan\c con is similar in spirit to the D6--brane case we study in
  detail first. Even if the D6--brane example we study here does not
  turn out to be salvageable in its entirety, we will see that the key
  features survive in these better--behaved examples.
\end{itemize}
In short, we can use the D6--brane case as a (rather pretty) simple
model, and be aware of the negative density, but not distracted by it,
leaving its understanding for another day.

We also exhibit the fact that the behaviour of the harmonic functions
corresponding to the distribution has a beautiful expansion in terms
of Legendre polynomials. Recalling that the moduli space of the
wrapped brane system is isomorphic to that of the Coulomb branch of
the ${\cal N}=4$ supersymmetric $(2+1)$--dimensional pure $SU(N)$
gauge theory, this will yield a useful geometrical parametrisation of
the vacuum expectation values of operators made from the symmetric
product of the three adjoint scalars in the gauge multiplet, an issue
we return to in section~\ref{gaugetheory}.

In section~\ref{wrapping} we uncover the properties of the wrapped
system, seeking the repulson and enhan\c con loci, and characterising
them, displaying the equations which result in figures~\ref{loci}
and~\ref{notsphere}. We probe the supergravity solution with wrapped
branes and point particles, in order to discover the nature of the
unphysical regions of the geometry. We then perform the excision to
construct new geometries which are free of the repulson regions
arising from the wrapping, checking consistency in supergravity in
section~\ref{excision}. In section \ref{others} we discuss some
features of wrapped D4-- and D5--brane distributions.  Section
\ref{gaugetheory} extracts some gauge theory results from the
D6--brane case. We conclude with some remarks in
section~\ref{conclude}.

\section{D--Brane Distributions}
\label{distribution}
It is possible to derive a metric for a continuous distribution of
branes by taking extremal limits of rotating solutions. The limits
remove the rotation and restore supersymmetry, and the parameters that
corresponded to rotation remain in the geometry as parameters of the
distribution (for D3--branes this was done first in
refs.\cite{klt,sfetsos-cft}).  We can do this here for D6--branes as
follows: The rotating black six--brane solution, in the usual
supergravity conventions, is given by\footnote{Solutions corresponding
  to rotating $p$--brane solutions of type~II supergravity have been
  found in refs.\cite{russo,klt,Harmark:1999xt} by uplifting of
  rotating black hole solutions of various types.}:
\begin{eqnarray}
ds^2 &=& \f_6^{-1/2} \Big( -K ~dt^2 + \sum_{i=1}^{6}dx_i^2 \Big) +
\f_6^{1/2} \Big( \bar{K}^{-1} \frac{\Delta}{\Xi}~dr^2 + \Delta r^2
~d\theta^2
+ \Xi r^2 \sin^2\theta ~d\phi^2 \Big) \nonumber \\
&& + \f_6^{-1/2} \frac{2m}{\Delta r} \Big( \ell^2 \sin^4\theta
~d\phi^2
- 2 \cosh\delta ~\ell \sin^2\theta ~dt d\phi \Big)\ , \nonumber \\
e^{2\Phi} &=& \f_6^{-3/2}, \nonumber \\
C_7 &=& \frac{1}{\sinh\delta} \f_6^{-1} \Big( \cosh\delta ~dt -
\ell \sin^2\theta ~d\phi \Big) \wedge dx_1 \wedge \ldots \wedge
dx_6\ ,
\labell{rotating}
\end{eqnarray}
where
\begin{equation}
\f_6 = 1 + \frac{2m \sinh^2\delta}{\Delta r}\ , \qquad \qquad K = 1 -
\frac{2m}{\Delta r}\ , \qquad \qquad \bar{K} = 1 - \frac{2m}{\Xi r}\ ,
\end{equation}
and
\begin{equation}
\Delta = 1 + \frac{\ell^2}{r^2}\cos^2\theta\ , \qquad \qquad \Xi = 1
+ \frac{\ell^2}{r^2}\ ,
\labell{params}
\end{equation}
while $r$ is a radial coordinate for three Cartesian coordinates
$x_7,x_8,x_9$ with $r^2=x_7^2+x_8^2+x_9^2$ and we are using standard
spherical polar coordinates such that $x_9=r\cos\theta$, and so on.

We can obtain an extremal limit by sending the non--extremality
parameter $m \rightarrow 0$ and the boost parameter $\delta
\rightarrow \infty$, while keeping $r_6 = 2m \sinh^2\delta$ fixed.  In
this limit the metric component, $g_{t\phi}$, giving rise to rotation
does not survive and the resulting solution is:
\begin{eqnarray} 
ds^2 &=& f_6^{-1/2} \Big( -dt^2 + \sum_{i=1}^{6}dx_i^2 \Big) +
f_6^{1/2} \Big( \frac{\Delta}{\Xi}~dr^2
+ \Delta r^2 ~d\theta^2 + \Xi r^2 \sin^2\theta ~d\phi^2 \Big)\ , \nonumber\\
e^{2\Phi} &=& f_6^{-3/2}, \nonumber\\
C_7 &=& f_6^{-1} dt \wedge dx_1 \wedge \ldots \wedge dx_6\ ,
 \labell{D6metric-oldcoordinates}
\end{eqnarray}
where
\begin{equation}
f_6 = 1 + \frac{r_6}{\Delta r}\ .
\end{equation}
This can be done for the other solutions presented in
refs.\cite{russo,klt,Harmark:1999xt} as well, giving distributions of
other D$p$--branes, and we refer to some of the results of this later.

The normalisation $r_6 = g N \alpha^{\prime 1/2}/2$ gives $N$ units of
D6--brane charge, $\mu_6=(2\pi)^{-6}\alpha^{'-7/2}$, in the standard
units~\cite{primer}.  In the next section we wrap this configuration
of D6--branes on K3. However, let us first examine the structure of this
configuration, in order to understand better what new features
are specifically brought about by the wrapping, and which are due to
the distribution's geometry.

The parameter $\ell$ controls the departure from a spherical geometry.
In the case $\ell=0$, the solution in
equation~(\ref{D6metric-oldcoordinates}) reduces to the usual
spherically symmetric static D6--brane solution\cite{horostrom}, where
the singularity at $r=0$ is interpreted as the position where the
branes reside. Now, as soon as $\ell \neq 0$, that singularity
disappears except for on the equatorial plane $\theta=\pi/2$: The
denominator in the harmonic function is now
$r+(\ell^2/r)\cos^2\theta$, which only vanishes at $r=0,\theta=\pi/2$,
so the source at $r=0$ has been modified. Let us try to understand
how.

To get a better understanding of the source, let us perform the
following transformation\cite{russo} to isotropic coordinates which
we shall refer  to as ``extended'':
\begin{eqnarray} 
y_1 & = & \sqrt{r^2 + \ell^2} \, \sin \theta \, \cos \phi\ , \nonumber \\
y_2 & = & \sqrt{r^2 +\ell^2} \, \sin \theta \, \sin \phi \ ,\nonumber \\
y_3 & = & r \, \cos \theta\ . \labell{extendedcoordinates}
\end{eqnarray}
We see that the origin $r=0$ is mapped to a disc, given by 
$y_1^2+y_2^2=\ell^2\sin^2\theta$, $y_3=0$. Going to $r=0$ along
$\theta=0$ takes us to the centre of the disc at $y=0$, while an
approach along $\theta=\pi/2$ takes us to the edge of the disc at
$y_1^2+y_2^2=\ell^2$. Approaching along other angles takes us to the 
interior of the disc.

The metric in equation~(\ref{D6metric-oldcoordinates})
reduces to a standard brane form:
\begin{equation}
ds^2 = F_6^{-1/2} \Big( -dt^2 + \sum_{i=1}^{6}dx_i^2 \Big) +
F_6^{1/2} \Big( dy_1^2 + dy_2^2 + dy_3^2 \Big)\ ,
\end{equation}
where the harmonic function is given by
\begin{equation}
F_6 = 1 + \frac{r_6 \sqrt{\Lambda + \Sigma}}{\sqrt{2}\Sigma}\ ,
\end{equation}
with
\begin{equation}
\Sigma = \sqrt{\Lambda^2 + 4 \ell^2 y_3^2}\ , \qquad
\Lambda = y^2 - \ell^2\ , \qquad
y = \sqrt{y_1^2 + y_2^2 + y_3^2}\ .
\end{equation} 
Now we see that the singularity we identified earlier is in fact a
{\it ring} of radius $\ell$. Is this where the branes are located?  The
harmonic function $F_6$ should have an integral representation
given schematically by
\begin{equation}
F_6 = 1 + \int{\frac{\sigma(y^{\prime})}{|y - y^{\prime}|} dy^{\prime}}\ ,
\end{equation}
for some density function $\sigma(y)$ representing a continuous
distribution of branes in the coordinate~$y$. How seriously we should
take the physical meaning of this distribution is a matter of
interest. For many applications, of concern to supergravity
quantities, the meaning of a continuous distribution of branes should
not be a problem, but we must remember that in the full string theory,
we might wish to probe the structure of the solution at resolutions
which might render the distribution meaningless.

It is possible to directly determine the distribution's dependence on
$y$, using an analogue of Gauss' Law. For six--branes, we have three
transverse spatial dimensions, and so the problem of determining
harmonic functions is in fact directly translated into an
undergraduate electromagnetism problem.

\begin{figure}[ht]
  \centering
  \scalebox{0.5}{\includegraphics{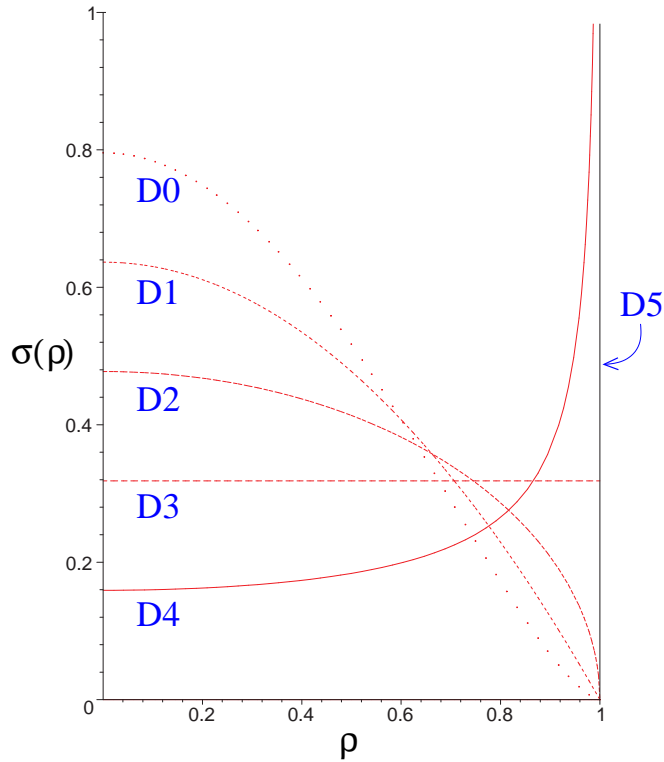}}
\caption{\footnotesize Disc 
  density distributions of D$p$--branes in a plane for $p=0,\cdots5$,
  normalised to the same area. The case of D5--branes is a delta
  function on the edge, at $\rho=1$. The amusing pattern followed by
  these distributions is described in the text.}
\label{distributions}
\end{figure}

Let us define $\rho = \sqrt{y_1^2 + y_2^2}$, $z = y_3$.  It
is sufficient to determine the harmonic function's behaviour along the
$z$--axis.  The density function and the angular
dependence of $F_6$ follows directly from harmonic analysis.
The analogue of Gauss' law in electrodynamics for a standard
infinitesimal ``pillbox'' surface defined on the $z=0$ plane is:
\begin{equation}
({\vec E}_+ - {\vec  E}_-) \cdot {\vec n} = 4 \pi \sigma\ ,
\end{equation}
where ${\vec n}$ is the unit normal vector of the surface directed
from one side ($-$) to the other side~($+$) of the surface, and
$\sigma$ is the surface charge density. The electric field is ${\vec
  E} = - {\vec \nabla} \Phi$, while the role of the potential $\Phi$
is now played by the harmonic function. Taking the derivative,
\begin{equation}
\frac{\partial F_6}{\partial z} =
\frac{z r_6}{\sqrt{2} \Sigma^3 \sqrt{\Lambda + \Sigma}}
\Big( \Sigma^2 - \Lambda \Sigma - 2 \Lambda^2
- 2 \ell^2 ( 2 \Lambda + \Sigma ) \Big)\ ,
\end{equation}
we obtain
\begin{eqnarray}
\sigma(\rho)  =  \frac{1}{4 \pi} \left(
\frac{\partial F_6}{\partial z} \Big|_{z \rightarrow 0^-}
- \frac{\partial F_6}{\partial z} \Big|_{z \rightarrow 0^+} \right) =  - \frac{r_6 \ell}{2 \pi (\ell^2 - \rho^2)^{3/2}}\ ,
\end{eqnarray}
where an expansion in  small $z$ was used to get this result.  This
density is negative, but happily (since it would be hard to see how to
get a positive result for the D6--brane charge from this), it
integrates to infinity over the disc, due to the boundary
contribution.  We must therefore add an extra positive term to the
boundary, so that the complete distribution is
\begin{equation}
\sigma(y) = - \frac{r_6\ell}{2 \pi 
(\ell^2 - \rho^2)^{3/2}} \Theta(\ell-\rho) \delta(z)
+ \frac{r_6}{2 \pi \sqrt{\ell^2 - \rho^2}} 
\delta(\ell-\rho) \delta(z)\ . \labell{D6density}
\end{equation}
It is worth checking that the normalisation of the configuration is
indeed $r_6$ as expected
\begin{eqnarray}
\int_0^{\ell} 2 \pi  \sigma(\rho) \rho d\rho & = &
\int_0^{\ell} \Big( - \frac{r_6 \ell \rho}{(\ell^2 - \rho^2)^{3/2}}
+ \frac{r_6 \rho}{\sqrt{\ell^2 - \rho^2}} \delta(\ell-\rho) \Big) d\rho \nonumber \\
& = &
\left[ - \frac{r_6 \ell}{\sqrt{\ell^2 - \rho^2}} \right]_{\rho=0}^{\rho=\ell}
+ \left.\frac{r_6 \rho}{\sqrt{\ell^2 - \rho^2}} \right|_{\rho=\ell}  = 
r_6\ ,
\end{eqnarray}
and that the brane distribution (\ref{D6density}) correctly reproduces
our harmonic function along the $z$--axis:
\begin{eqnarray}
F_6 & = &
1 + 2 \pi \int_0^{\ell} \frac{ \sigma(\rho)\rho d\rho}{\sqrt{z^2+\rho^2}}  \nonumber \\
& = &
1 + \int_0^{\ell} \left( - \frac{r_6 \ell \rho }{\sqrt{z^2+\rho^2} (\ell^2-\rho^2)^{3/2}}
+ \frac{r_6 \rho}{\sqrt{z^2+\rho^2}\sqrt{\ell^2 - \rho^2}} \delta(\ell-\rho) \right) d\rho \nonumber \\
& = &
1 + \left[ - \frac{r_6 \ell}{(\ell^2 + z^2)} \frac{\sqrt{z^2 + \rho^2}}{\sqrt{\ell^2-\rho^2}}
\right]_{\rho=0}^{\rho=\ell}
+ \left.\frac{r_6 \rho}{\sqrt{z^2+\rho^2}\sqrt{\ell^2 - \rho^2}} \right|_{\rho=\ell} 
 = 
1 + \frac{r_6 z}{\ell^2 + z^2}\ .\labell{theform}
\end{eqnarray}
This is very interesting, particularly when compared to the results
(listed in equation~\reef{more}) one can get by computing the
analogous quantities for D$p$--brane disc distributions for
$p=0,\cdots5$. The results are easy to obtain\footnote{The case of
  disc distributions of D3--branes was known before, being dual to
  part of the Coulomb branch of the ${\cal N}=4$ $SU(N)$ gauge
  theory\cite{klt,sfetsos-cft}.} by noting first that along the $z$
axis, the harmonic function's form in extended coordinates is exactly
the same as in unextended coordinates:
\begin{displaymath}
  f=1+\frac{r_p^{7-p}}{z^{7-p}\left(1+{\ell^2}/{z^2}\right)}\ .
\end{displaymath}
This can be verified by simply examining the appropriate metrics,
which can be derived by taking an extremal limit of the metrics listed
in \cite{Harmark:1999xt}, as we did for equation~\reef{rotating} to
get equation~\reef{D6metric-oldcoordinates}, keeping only one non--zero 
parameter $\ell_1 = \ell$. Given that the harmonic
function should have an integral representation
\begin{displaymath}
  f=1+\int \frac{\sigma(y^\prime)dy^\prime}{|y-y^\prime|^{7-p}}\ ,
\end{displaymath}
it is easy to guess the densities in each case and check them by explicit
integration.  We list the densities here, and plot them in
figure~\ref{distributions}:
\begin{eqnarray}
{\rm D0}: & & \sigma(y) \ = \ +\frac{5r_0^7}{2\pi \ell^5}\ (\ell^2 - \rho^2)^{\frac{3}{2}} \ \Theta \,(\ell-\rho)\ \delta(z)\ ,\nonumber \\
{\rm D1}: & & \sigma(y) \ = \ +\frac{4r_1^6}{2\pi \ell^4}\ (\ell^2 - \rho^2)^{1}\ \Theta \,(\ell-\rho) \ \delta(z)\ ,\nonumber \\
{\rm D2}: & & \sigma(y) \ = \ +\frac{3r_2^5}{2\pi \ell^3}\ (\ell^2 - \rho^2)^{\frac{1}{2}} \ \Theta \,(\ell-\rho)\ \delta(z)\ ,\nonumber \\
{\rm D3}: & & \sigma(y) \ = \ +\frac{2r_3^4}{2\pi \ell^2}\ (\ell^2 - \rho^2)^{0} \ \Theta \,(\ell-\rho)\ \delta(z)\ ,\nonumber \\
{\rm D4}: & & \sigma(y) \ = \ +\frac{r_4^3}{2\pi \ell}\ (\ell^2 - \rho^2)^{-\frac{1}{2}} \ \Theta \,(\ell-\rho)\ \delta(z)\ ,\nonumber \\
{\rm D5}: & & \sigma(y) \ = \ \hskip 5.35cm+\frac{r_5^2}{2\pi }\ (\ell^2 - \rho^2)^{0}\delta(\ell - \rho)\ \delta(z)\ ,\nonumber \\
{\rm D6}: & &
\sigma(y) \ = \ - \frac{r_6\ell}{2 \pi} (\ell^2 - \rho^2)^{-{3\over2}} \Theta(\ell-\rho) \delta(z)
+ \frac{r_6}{2 \pi}(\ell^2 - \rho^2)^{-{1\over2}} \delta(\ell-\rho) \delta(z)\ .\labell{more}
\end{eqnarray}
(Now $z$ in the above is shorthand for all of the directions
transverse to the disc and, of course, to the brane's world--volume.)
The pattern is amusing. For a D$p$--brane, as $p$ becomes larger, the
distribution increasingly spreads away from the centre.  The D5--brane
case is a limiting one, having all of the branes at the boundary
forming a ring.  The D6--brane case (not plotted) also has a
$\delta$--function on the boundary, but is accompanied by a negative
contribution to $\sigma(y)$ in the interior.

While there is a striking pattern here, we must pause to consider what
the physical significance of the negative contribution to the
distribution of branes might be\footnote{In fact, this has occurred
  previously in a related context in ref.\cite{freed}, where
  appropriately cautious statements were made. The context was D3--brane
  distributions (in the decoupling limit), describing part of the
  Coulomb branch of the dual ${\cal N}=4$ $SU(N)$ gauge theory.
  There, switching on a vacuum expectation value of a perfectly
  physical operator in the gauge theory corresponded to a five
  dimensional ball distribution, which by happy coincidence is
  described by the same function as we have listed above for the
  D6--brane disc.}. As we said previously, we must be careful about
the meaning of the continuous distribution in general, since it is
merely a supergravity approximation. However, the negative density (and
hence tension and charge) is somewhat harder to accept, and  may be
another cry for help from the supergravity, appealing to more stringy
physics to resolve the problem. 

This is of course reminiscent of the features which are resolved by
the enhan\c con mechanism. There is a region where branes seem to have
negative tension. In that case\cite{jpp} this was seen by sending in a
probe brane of the same type to the affected region. The supergravity
geometry in the interior was consequently discarded because one could
argue that it can not be constructed by bringing all of the constituent
branes from infinity. We can {\it not} apply that reasoning here,
however. A probe D6--brane will again see no potential for movement in
this background, and the resulting moduli space metric is perfectly
flat right down to $y=0$.  So it would seem that we can not argue in
this way that we can't construct this supergravity solution.

Later, we will also see that this sort of distribution corresponds to
switching on apparently sensible vacuum expectation values for gauge
theory operators.  That discussion will be aided by noting that the
harmonic function $F_6$ can be expanded in a series in the variable
${\ell}/{y}$ for $y>\ell$, or ${y}/{\ell}$ for $y<\ell$. It is
delightful to see how these series arrange themselves in terms of
Legendre polynomials:
\begin{eqnarray}
F_6 & = & 1 + \frac{r_6}{y} \sum_{n=0}^{\infty} (-1)^n
\left(\frac{\ell}{y}\right)^{2n} P_{2n}(\cos\ \hat{\theta})\ ,
\qquad y > \ell\ , \\
F_6 & = & 1 + \frac{r_6}{\ell} \sum_{n=0}^{\infty} (-1)^{n}
\left(\frac{y}{\ell}\right)^{2n+1} P_{2n+1}(\cos\ \hat{\theta})\ ,
\qquad y < \ell\ ,
\labell{Legendre}
\end{eqnarray}
where we have defined new polar coordinate angles with respect to the
$y_i$'s:
\begin{eqnarray}
  \hat{\theta}&=&\cos^{-1}\left(\frac{y_3}{ y}\right)\ ,\nonumber\\
\hat{\phi}&=&\tan^{-1}\left(\frac{y_2}{ y_1}\right)=\phi\ ,\labell{newangles}
\end{eqnarray}
and $P_n(x)$'s are the Legendre polynomials in the variable $x$, with
the normalisation:
\begin{equation}
P_0(x)=1\ ; 
\quad P_1(x)=x\ ; \quad P_2(x)=\frac{1}{2}(3x^2-1)\ ;\quad{\rm\etc.}
\labell{normalised}
\end{equation}
In hindsight, this is of course what we should expect from a direct
expansion of equation~\reef{theform}, combined with 
harmonic analysis.

It is worth noting that we can insert any of our favourite harmonic
potentials from electromagnetism (or higher dimensional
generalisations) and get a supergravity solution corresponding to a
distribution of D--branes. This could in principle be wrapped on K3,
as we do in section~\ref{wrapping}.  However, since these are most
commonly found as a series expansion of the form above, exact
determination of the crucial enhan\c con and repulson loci (as we do
later) will not in general be possible.

\subsection{Particle Probes of the Geometry} 
\label{probing}
\begin{figure}[ht]
  \centering
  \scalebox{0.7}{\includegraphics{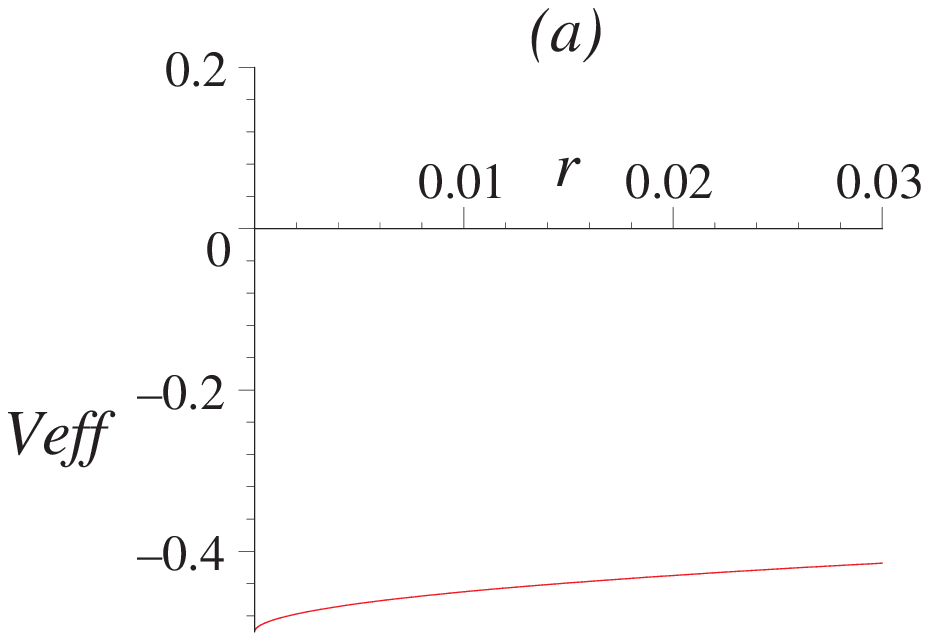}}
  \hskip1cm\scalebox{0.7}{\includegraphics{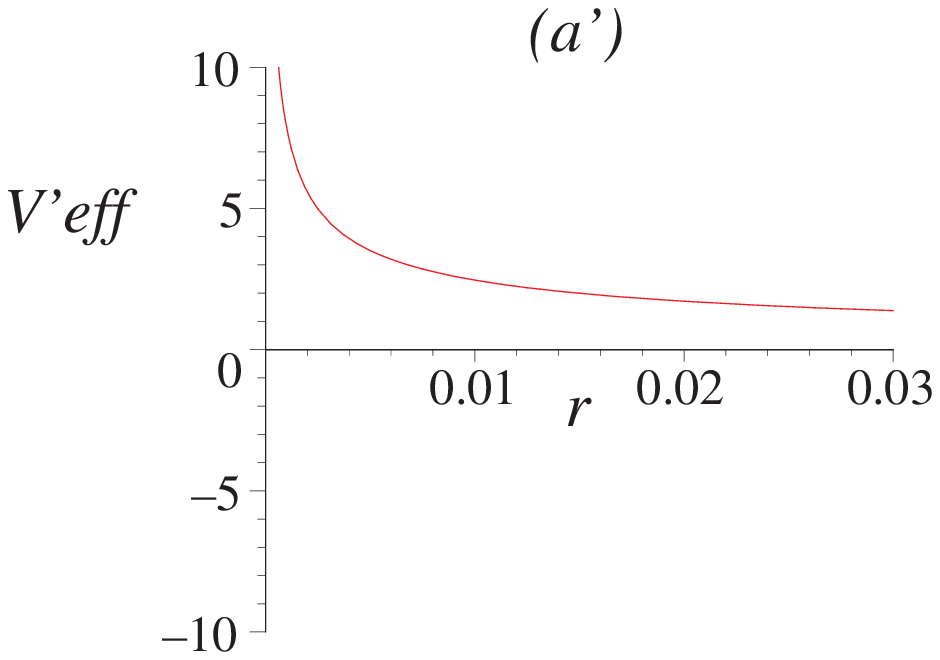}}
  \\ \vspace{2mm}
\scalebox{0.7}{\includegraphics{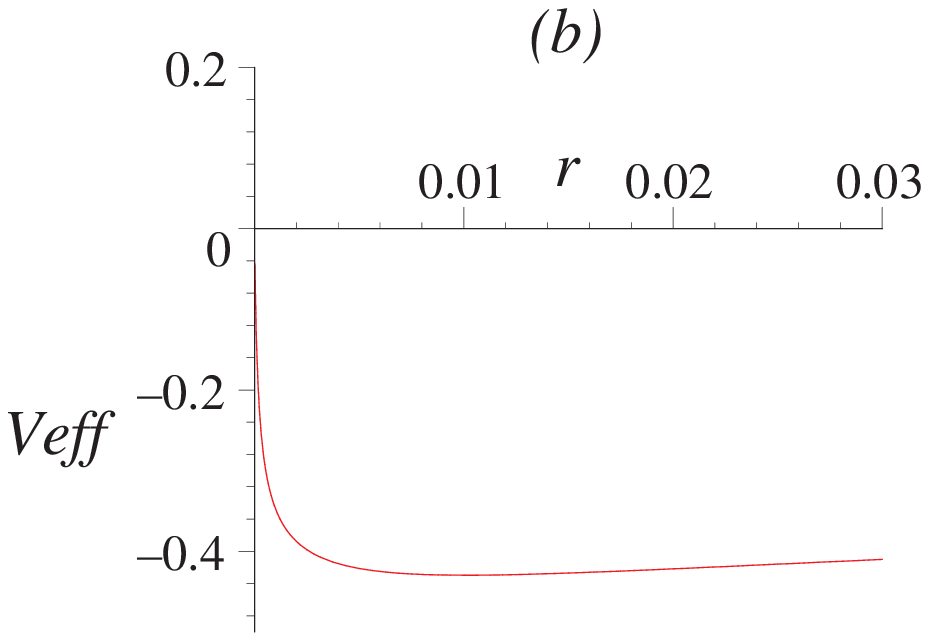}}
  \hskip1cm\scalebox{0.7}{\includegraphics{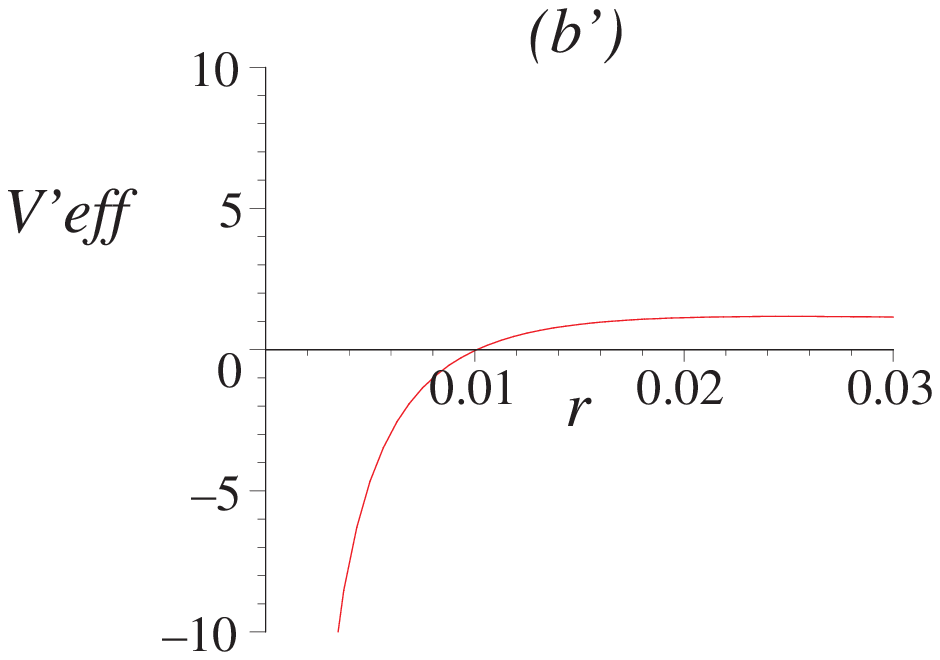}}
  \caption{\footnotesize
    Gravitational features of the unwrapped D6--brane distribution as
    seen by a neutral test particle.  The test particle effective
    potential (left) and its derivative (right) along the axis of
    symmetry ($\theta=0$) corresponding to {\it (a)} $\ell=0$, {\it
      (b)} $\ell > 0$. In the latter case the potential becomes
    repulsive at a distance $\ell$ from the origin.  The effective
    potential for a particle moving on the equatorial plane
    ($\theta=\pi/2$) is always attractive, and is similar to case {\it
      (a)}.  }
  \label{Veffa}
\end{figure}

It is interesting to probe the system with a point particle using the
standard technology.  There are Killing vectors
$\mathbf{\xi}=\partial/\partial t$ and ${\bf \eta}=\partial/\partial
\phi$ which, for a particle moving on timelike geodesics, with velocity
${\bf u}$, define for us conserved quantities $e
= - {\bf \xi} \cdot {\bf u}$ and $L = {\bf \eta} \cdot {\bf u}$.
In terms of these, we can write a first integral of the geodesic
equation.  The purely radial motion on the equatorial plane or along
the symmetry axis (\ie, assuming ${\bf u}^{\theta}=0$, and ${\bf
  u}^{\phi}=0$) of the test particle is described by
\begin{equation}
\frac{\Delta}{\Xi} \dot{r}^2 = E - V_{\rm eff}(r)\ .
\end{equation}
The effective potential that the particle probe feels is
\begin{equation}
V_{\rm eff}(r)= - \frac{1}{2}\left(g_{tt} + 1\right)\ .
\labell{veef}
\end{equation}
We plot $V_{\rm eff}$ for a particle approaching along the equator and
also along the $z$--axis in figure~\ref{Veffa}.  It is interesting to
note that accompanying this negative brane region is a repulsive
behaviour in the supergravity. A test particle located along the
$z$--axis sees a potential which is repulsive inside a radius $\ell$.
This has nothing to do with the repulson singularity coming from
wrapping, which we will study shortly, so we must bear in mind that it
is a feature which will descend from the unwrapped configuration to
the wrapped configuration. Later, in section~\ref{others}, we shall
see that there is no such behaviour for D5--brane distributions, and
so we will treat this as an artifact of the D6--brane case in what
follows, as it will not affect our analysis of the excision of the
repulson behaviour arising from wrapping.

\section{Wrapping the D6--Branes}
\label{wrapping}
We have learned enough about our configuration to return to our
problem of wrapping the branes on K3.  Wrapping a D6--brane on K3
induces precisely one unit of negative D2 brane charge in the
D6--brane world--volume. As in ref.\cite{jpp}, this suggests that the
appropriate supergravity solution is that appropriate to  a D6--brane
with delocalised D2--branes in its world--volume\cite{tseytlin}, where
in the D2--brane harmonic function, we put $r_2 = - r_6 V_{\star}/V$,
where $V$ is the volume of K3 as measured at infinity and $V_{\star} =
(2\pi)^4 \alpha^{\prime 2}$.  This leads to the following geometry (we
remind the reader here that there are no real D2--branes in the
geometry):
\begin{eqnarray} 
ds^2 &=& f_2^{-1/2}f_6^{-1/2}( - dt^2 + dx_1^2 + dx_2^2)
+ f_2^{1/2}f_6^{1/2}
\left(\frac{\Delta}{\Xi}\, dr^2 + \Delta r^2 \,d\theta^2 + \Xi \,r^2
 \sin^2 \theta \,d\phi^2\right)
\nonumber\\
&&\hskip3cm 
 + V^{1/2}f_2^{1/2} f_6^{-1/2}
 ds_{K3}^2\ .\labell{D6D2metric-oldcoordinates}
\end{eqnarray}
with
\begin{equation}
f_2 = 1 - \frac{r_6 V_{\star}}{V \Delta r}\ , \quad
f_6 = 1 + \frac{r_6}{\Delta r}\ ,
\end{equation}
$ds_{K3}^2$ is the metric of a unit volume K3 surface, and the
functions $\Delta$ and $\Xi$ were defined before.  The dilaton and
R--R fields are given by
\begin{eqnarray}
e^{-2\Phi} & = & \left(\frac{f_6^3}{f_2}\right)^{1/2}\ , \nonumber\\
C^{(3)} & = & \frac{1}{f_2} dt \wedge dx_1 \wedge dx_2\ ,\nonumber\\
C^{(7)} & = & \frac{1}{f_6} dt \wedge dx_1 \wedge dx_2 \wedge V ds_{K3}\ .
\end{eqnarray}
Despite being a solution to the supergravity equations of motion, the
geometry for this configuration is not consistent. Naked singularities
(seen \eg\ by examining the curvature invariants $R$,
$R_{\mu\nu}R^{\mu\nu}$, $R_{\mu\nu\rho\sigma}R^{\mu\nu\rho\sigma}$,
\etc.) of repulson\cite{repulson} type appear where the running K3
volume, given by:
 \begin{equation}
V(r,\theta)=V \frac{f_2}{f_6}\ ,\labell{k3volume}
\end{equation}
shrinks to zero. Some algebra shows that this occurs  at radii:
\begin{equation} 
r_{\rm r} = \frac{r_6 V_{\star}}{2 V} \left[ 1 \pm
\left( 1 - \frac{4 V^2}{r_{6}^{2} V_{\star}^{2}} \ell^2 \cos^2 \theta \right)^{1/2} \right].\labell{Singularityradius}
\end{equation}
When $\ell$ is zero, we have the spherically symmetric
situation where the singularity appears on a sphere of radius
${r_6 V_{\star}}/{V}$. For non--zero, but sufficiently small
$\ell$ the singularity appears at two disconnected loci,
one of them inside the other, and  between these loci
the metric (\ref{D6D2metric-oldcoordinates}) is imaginary.
When $\ell$ reaches the critical value
\begin{equation}
\ell^{\rm cr}_{\rm r} = \frac{r_6 V_\star}{2 V}\ ,
\end{equation}
these two surfaces meet and join into one single
surface for $\ell > \ell^{\rm cr}_{\rm r}$. 

We expect stringy effects to have switched on long before a vanishing
volume is reached, since when the volume gets to the value
$V_{\star}$, there are extra massless states coming from wrapped D4--
and anti--D4--branes, giving an enhanced $SU(2)$ in spacetime. The
radius at which this occurs is the enhan\c con radius, and it is
easily computed to give:
\begin{equation} 
r_{\rm e} = \frac{r_6 V_\star}{V - V_\star} \left[ 1 \pm \left( 1 - \frac{(V
 - V_\star)^2}{r_6^2V_\star^2} \ell^2 \cos^2 \theta \right)^{1/2} \right]\ .\labell{Enhanconradius}
\end{equation}

This is of course the same radius that gives a zero of the effective
tension of a probe wrapped D6--brane, whose action is\cite{jpp,primer}:
\begin{equation}
S = - \int d^3 \xi \,\,\mu_6  e^{-\Phi}  \left( V(r)-V_{\star} \right)
(- \det g_{ab})^{1/2} + \mu_6 \int C^{(7)} - \mu_2 \int C^{(3)}\ ,
\end{equation}
where $\xi^a$ $(a,b=0,1,2)$ are the coordinates on the unwrapped part of
the world--volume and $g_{ab}$ is the induced metric.
Working in static gauge the potential vanishes and we obtain the
 result for the effective Lagrangian:
\begin{equation}
{\cal L} =   \frac{\mu_6}{2g_s} (V f_2 - V_{\star} f_6) \,
\left(\frac{\Delta}{\Xi} \ \dot{r^2} + \Delta r^2 \ \dot{\theta^2}
+ \Xi\,\, r^2 \sin^2\theta \ \dot{\phi}^2\right)\ ,
\end{equation}
and we can read off the effective mass (tension) as
\begin{equation}
\tau_{\rm  eff}=\frac{\mu_6}{g_s}  f_6 \left( V \frac{f_2}{f_6} - V_{\star}\right)\ .
\end{equation}
Let us study the enhan\c con radius given in
equation~\reef{Enhanconradius}.  For $\ell=0$ we recover the
spherically symmetric case with enhan\c{c}on radius ${2 r_6
  V_\star}/({V - V_\star})$. For non--zero $\ell$, two different
situations can be observed, depending on whether $\ell$ is smaller or
greater than the critical value:
\begin{equation}
\ell_{\rm e}^{\rm cr} \ = \ \frac{r_6 V_\star}{V - V_\star}\ .
\end{equation}

When $\ell \leq \ell_{\rm e}^{\rm cr}$, there are two enhan\c con
shells which divide our geometry into three distinct regions.  The
tension of a D6--brane probe drops to zero at the outer enhan\c con
shell.  Let us call the exterior region of positive tension region I.
In between the outer and inner enhan\c con shells, the tension of our
probe would be negative.  We will call this region II.  This is the
region where we encountered the repulson
singularities~\reef{Singularityradius} and it will be excised shortly.
Finally, the tension is again positive in region III, inside the inner
enhan\c con shell. We can be sure that singularities are contained in
the region II only, because the running K3 volume is a continuous
function and $V_{\star}>0$. The origin $r=0$, appears to be
problematic, since it solves both equation~(\ref{Singularityradius})
and equation~(\ref{Enhanconradius}).  In fact, if we approach $r=0$
along the $z$--axis, we get $V(r,\theta)\to V$ while an approach along
the equator shows that $V(r,\theta)\to -V_{\star}$, which is puzzling.
This is partly resolved by going to the extended coordinates, where
$r=0$ opens up into a disc. Then we see that $V(r,\theta)\to
-V_{\star}$ as we approach the edge of the disc, while $V(r,\theta)=
V$ on the interior of the disc.

Notice what happens when $\ell = \ell_{\rm e}^{\rm cr}$.  The inner
and outer enhan\c con shells meet at two points on the $z$--axis.  The
tension of a brane probe moving along  this axis drops to zero at
the enhan\c con radius and becomes positive again.

For $\ell > \ell_{\rm e}^{\rm cr}$, the inner and outer enhan\c con
shells have merged into a single connected shell with a toroidal
shape.  Our geometry is now divided into two distinct regions.  The
volume of K3 is less than $V_{\star}$ in the interior of the torus.
The repulson singularities lie inside the torus. Actually, although
there is no physical significance to the fact, it is worth noting that
the repulson loci undergo a similar evolution from disconnected to
connected (with the critical value $\ell_{\rm r}^{\rm cr}$ separating
the two cases), with shapes of the same sort. Since $\ell_{\rm r}^{\rm
  cr}<\ell_{\rm e}^{\rm cr}$, the repulson always becomes connected
before the enhan\c con locus does.

\begin{figure}[ht]
  \centering
  \includegraphics{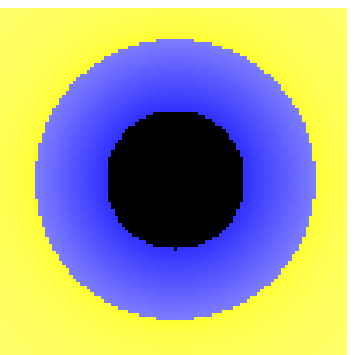}
  \includegraphics{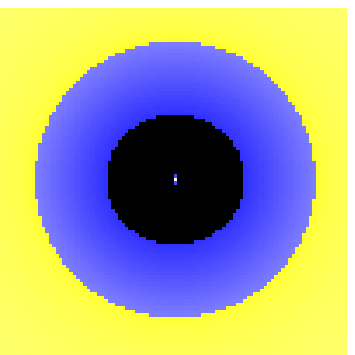}
  \includegraphics{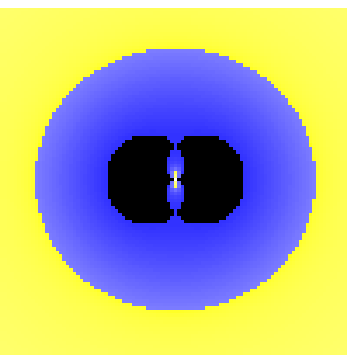}
  \includegraphics{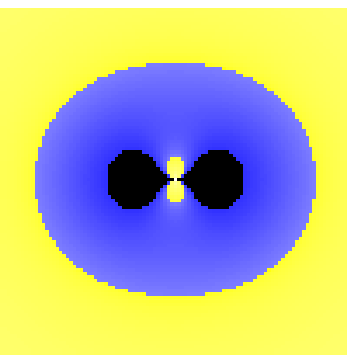}
  \\ \vspace{2mm}
  \includegraphics{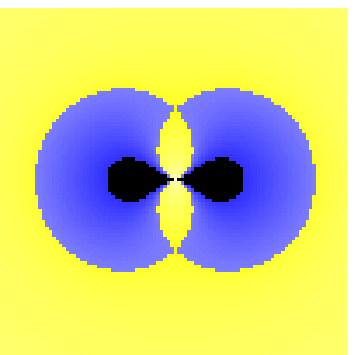}
  \includegraphics{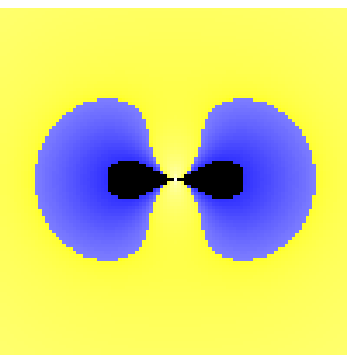}
  \includegraphics{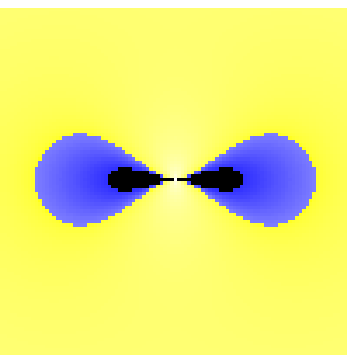}
  \includegraphics{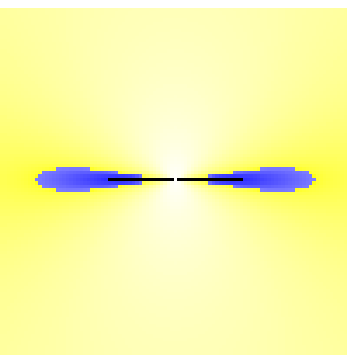}
  \caption{\footnotesize 
    A vertical slice through the non--spherical wrapped D6--brane
    geometry before excision.  The origin $r=0$ is placed in the
    centre of the figure. The symmetry axis, $\theta=0$, 
    corresponds to the vertical direction and $\theta=\pm {\pi}/{2}$ 
    to the horizontal.
    The region where the K3 volume $V(r,\theta)<0$ is
    black.  The region where $0 \leq V(r,\theta) < V_{\star}$ is dark
    grey (blue).  The repulson radius $r_{\rm r}$ is marked by the
    border between these two regions. The physically acceptable
    region $V(r,\theta) \geq V_{\star}$, where the tension of the
    probe brane is positive, is in the light shade (different shades
    between yellow ($V(r,\theta)$ is close to $V_{\star}$) and white
    ($V(r,\theta)$ is close to $V$)). The enhan\c con radius $r_{\rm
      e}$ is the border between the dark grey (blue) and the light
    regions. $\ell = 0$ gives a spherical enhan\c con.  
    For increasing values of $\ell$, two disconnected shells appear 
    forming a double enhan\c con.  Once $\ell^{\rm cr}_{\rm e}$ is
    exceeded, the two shells join forming a single enhan\c con. 
    This pattern can be seen explicitly in the figure for the following
    values of $\ell$ :\  
    $\ell = 0$ (top, left), $\ell = \frac{1}{2}\ell^{\rm cr}_{\rm r}$,
    $\ell = \ell^{\rm cr}_{\rm r}$, $\ell = \ell^{\rm cr}_{\rm
      e}-\frac{1}{2}(\ell^{\rm cr}_{\rm e}-\ell^{\rm cr}_{\rm r})$
    (top, right), $\ell = \ell^{\rm cr}_{\rm e}$ (bottom, left), $\ell
    = \frac{9}{8}\ell^{\rm cr}_{\rm e}$, $\ell = 2\ell^{\rm cr}_{\rm
      e}$, $\ell = 8\ell^{\rm cr}_{\rm e}$ (bottom, right).  Other
    parameters $r_6$, $V$, and $V_{\star}$ have been fixed to some
    typical values.}
  \label{K3volume-oldcoordinates}
\end{figure}

This all seems rather complicated, but is quite beautiful to look at,
in fact. We have already sketched the enhan\c con loci in
figures~\ref{loci} and~\ref{notsphere}, but now we put it together
with the repulson loci by shading in dark grey (or blue for viewers in
colour) in the regions where the volume is greater than zero but less
than $V_{\star}$ in figure~\ref{K3volume-oldcoordinates}, for
different values of $\ell$.  The light (or yellow) regions are volumes
greater than~$V_{\star}$, and so the boundary between these is the
enhan\c con locus. The black region is for negative volumes, in other
words, where the metric does not exist.  The boundary between it and
the grey (blue) is the repulson locus.

\begin{figure}[ht]
  \centering
  \includegraphics{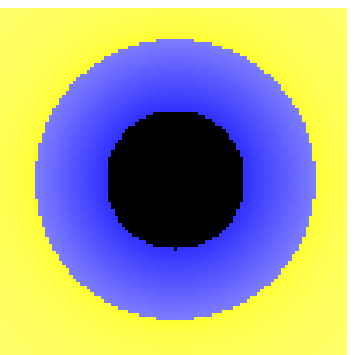}
  \includegraphics{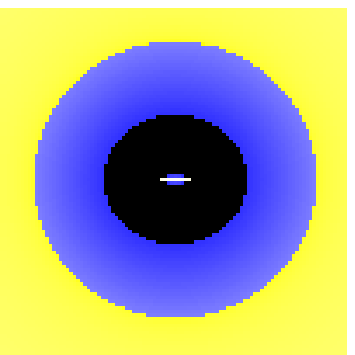}
  \includegraphics{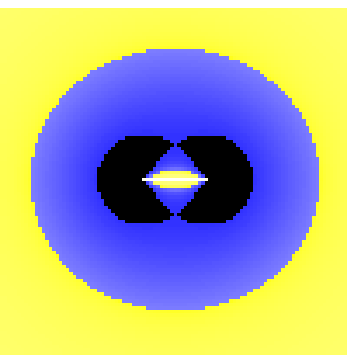}
  \includegraphics{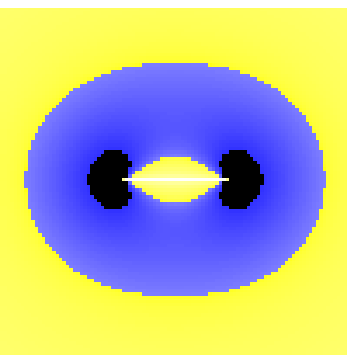}
  \\ \vspace{2mm}
  \includegraphics{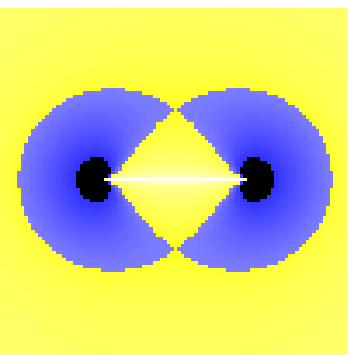}
  \includegraphics{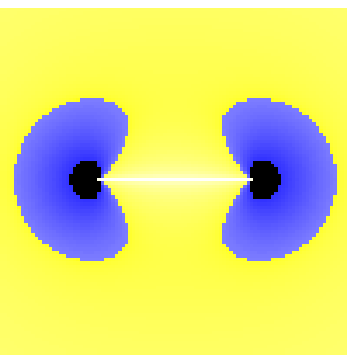}
  \includegraphics{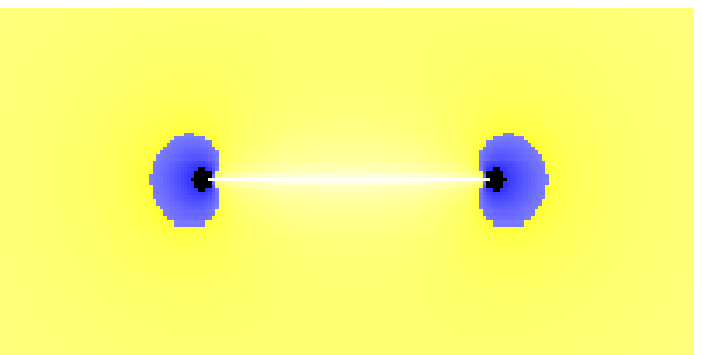}
  \caption{\footnotesize 
    A slice through the untreated wrapped D6--brane geometry in
    ``extended'' coordinates.  The colour coding is the same as in
    figure~\ref{K3volume-oldcoordinates}. The case $\ell = 0$ gives a
    spherical enhan\c con. For increasing values of $\ell$, two
    disconnected enhan\c con shells appear forming coconut shape.
    Once $\ell^{\rm cr}_{\rm e}$ is exceeded, the two enhan\c con
    shells join to form a torus.  This pattern can be seen for the
    following values of $\ell$ : \ $\ell = 0$ (top, left), $\ell =
    \frac{1}{2}\ell^{\rm cr}_{\rm r}$, $\ell = \ell^{\rm cr}_{\rm r}$,
    $\ell = \ell^{\rm cr}_{\rm e}-\frac{1}{2}(\ell^{\rm cr}_{\rm
      e}-\ell^{\rm cr}_{\rm r})$ (top, right), $\ell = \ell^{\rm
      cr}_{\rm e}$ (bottom, left), $\ell = \frac{9}{8}\ell^{\rm
      cr}_{\rm e}$, $\ell = 2\ell^{\rm cr}_{\rm e}$ (bottom, right).
    }
  \label{K3volume-extendedcoordinates}
\end{figure}

One may wonder what happens to the enhan\c{c}on if we transform the
solution (\ref{D6D2metric-oldcoordinates}) into the ``extended''
coordinates given in equations~(\ref{extendedcoordinates}). 
In these coordinates, essentially the same features arise. The
singular region where the K3 volume shrinks to zero is surrounded by
the enhan\c{c}on shell(s).  The same critical value of the
parameter~$\ell$ separates two families of enhan\c{c}ons as before. We
re--plot the K3 volume in the extended coordinates in
figure~\ref{K3volume-extendedcoordinates}.

\subsection{Probing the  Geometry} 
\label{Repulson-geometry-section}
Let us probe the system with a point particle, as done previously for
the purely spherical case\cite{jmpr}, and above in
subsection~\ref{probing} for the unwrapped case. The analysis is the
same, giving the effective potential~\reef{veef}, as before, but now
we insert the metric components for the putative wrapped geometry.
The effective potential is singular (\ie, exhibits infinite repulson)
when $g_{tt}$ is singular, and this happens at the repulson radius
$r_{\rm r}$.  The border between repulsive and attractive regions
corresponds to the minimum of the potential, which occurs when
\begin{equation} 
\frac{\partial}{\partial r} \ (f_2 f_6) = 0 \ .
\labell{repulsive-potential-minimal}
\end{equation}
It is significant that the relation
(\ref{repulsive-potential-minimal}) is satisfied exactly at $r=r_{\rm
  e}$, \ie, on the enhan\c{c}on shell.

\begin{figure}
  \centering
  \scalebox{0.7}{\includegraphics{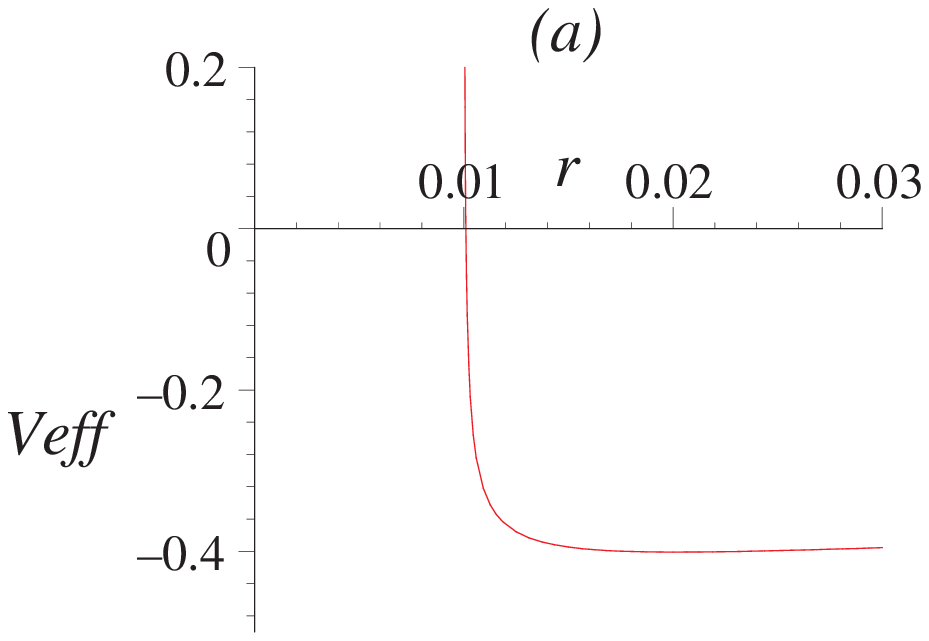}}
  \hskip1cm\scalebox{0.7}{\includegraphics{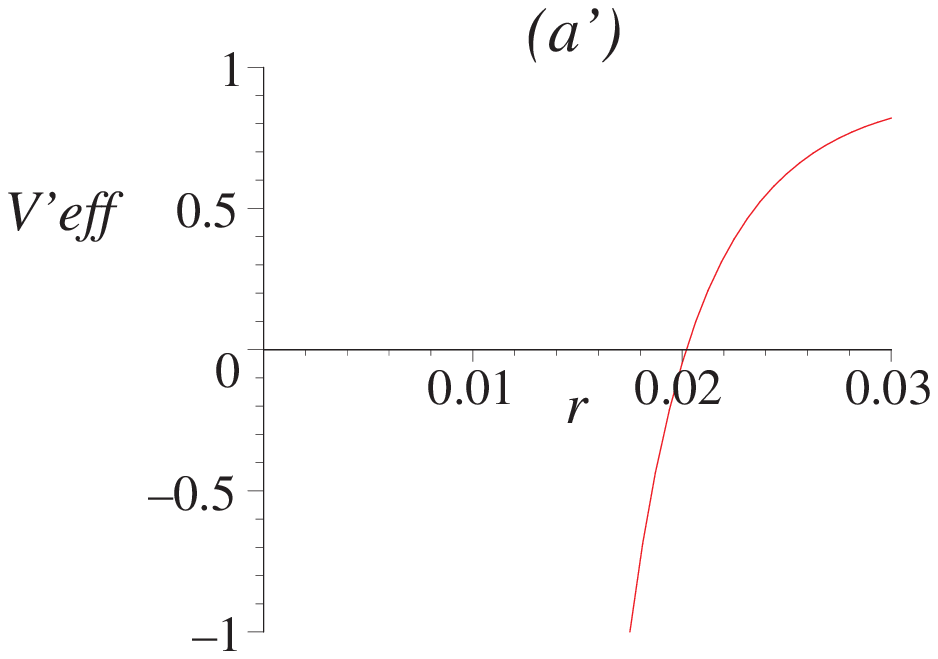}}
  \scalebox{0.7}{\includegraphics{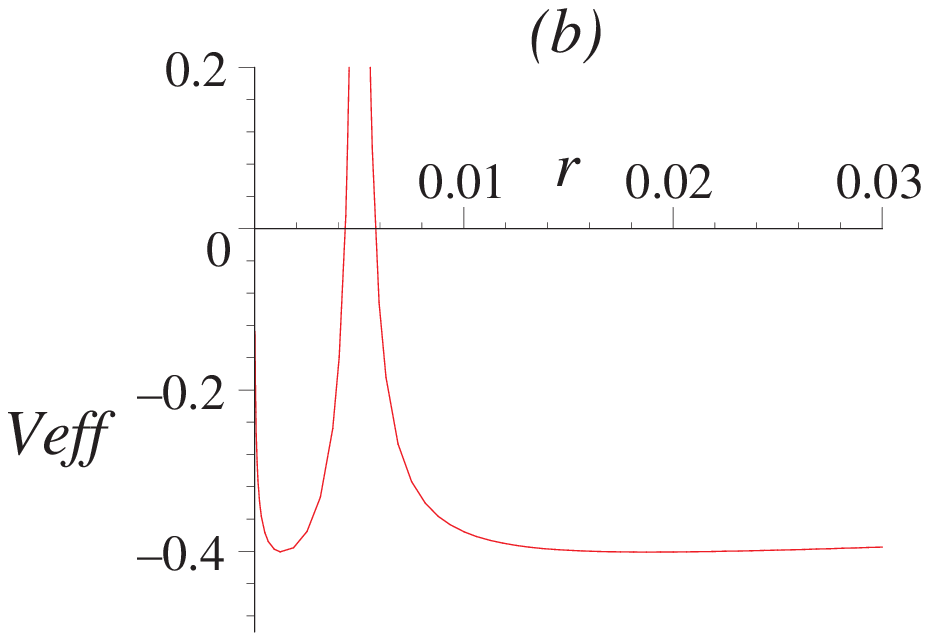}}
  \hskip1cm\scalebox{0.7}{\includegraphics{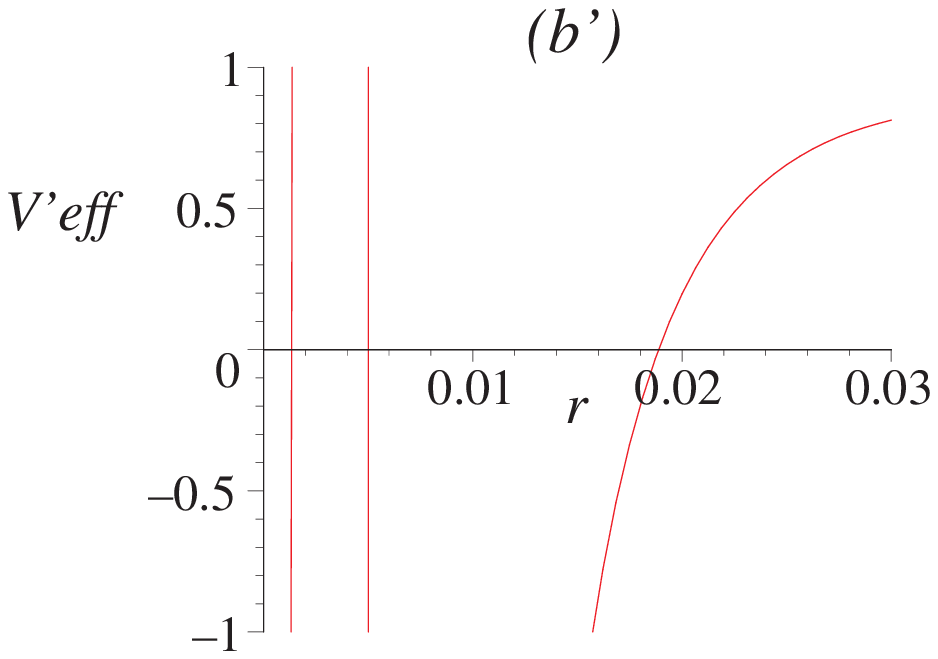}}
 \\ \vspace{2mm}
  \scalebox{0.7}{\includegraphics{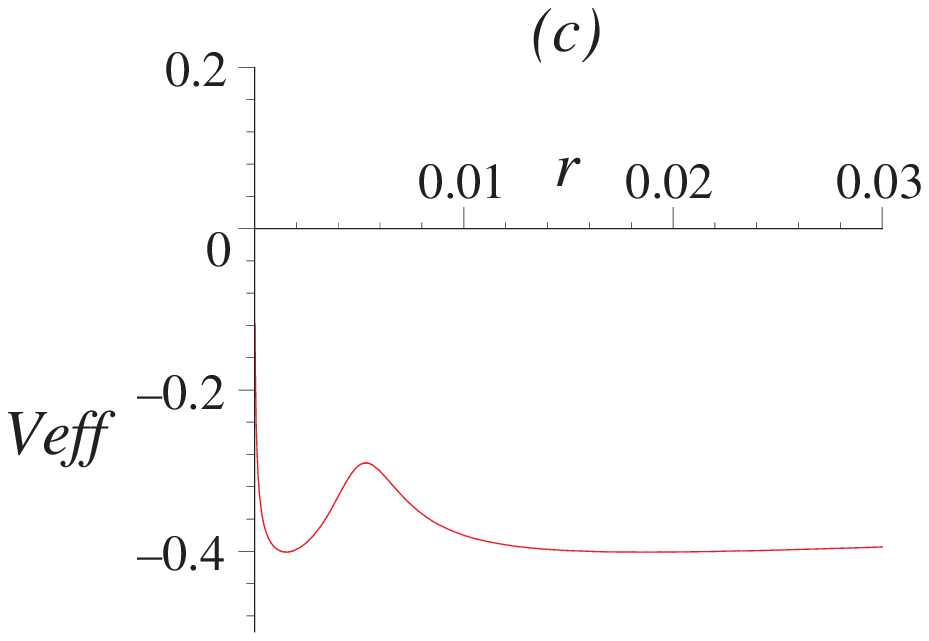}} 
  \hskip1cm\scalebox{0.7}{\includegraphics{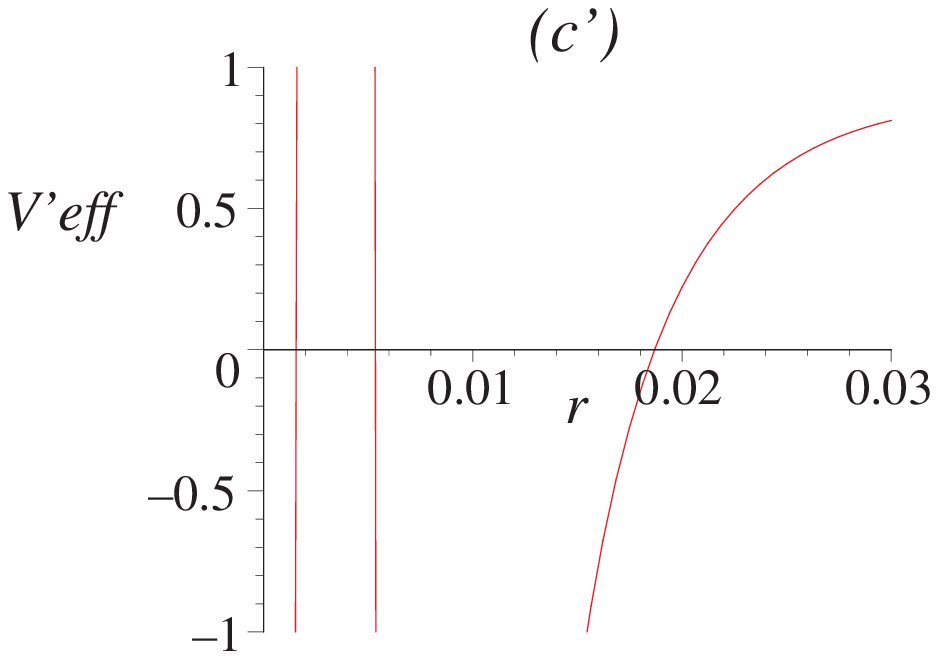}}
  \scalebox{0.7}{\includegraphics{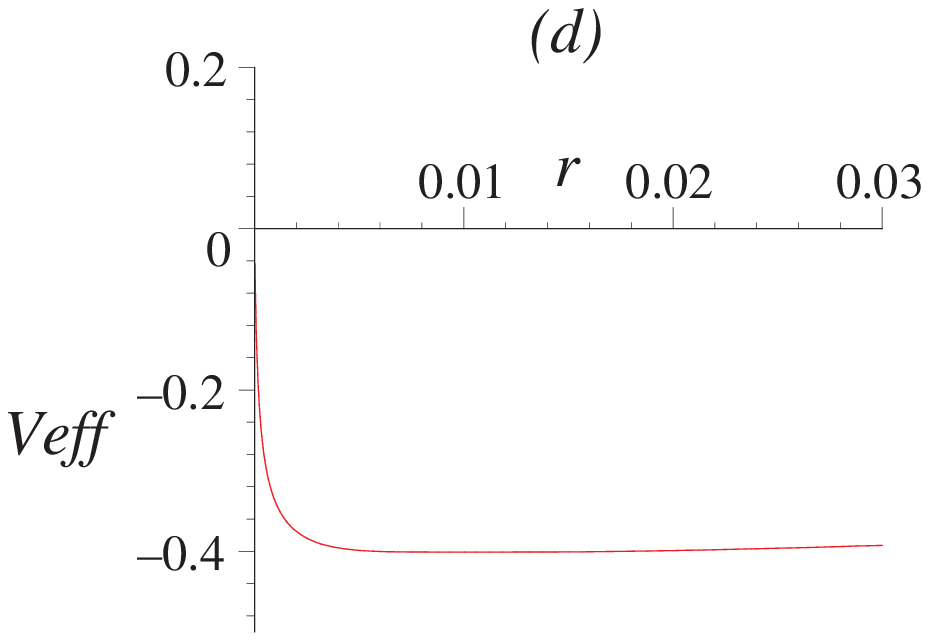}}
  \hskip1cm\scalebox{0.7}{\includegraphics{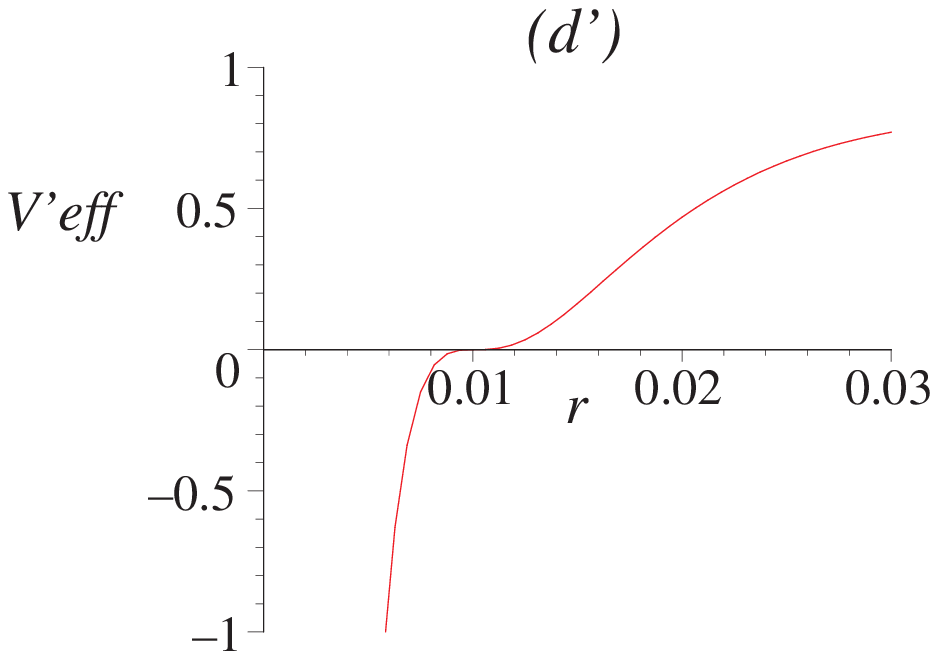}}
  \caption{\footnotesize
    Gravitational features of the D6--brane distribution wrapped on
    K3 as seen by a neutral test particle.  Test particle effective
    potential (left) and its derivative (right) along the axis of
    symmetry ($\theta=0$) corresponding to different values of
    parameter $\ell$: {\it (a)} $\ell=0$, {\it (b)} $\ell=\ell^{\rm
      cr}_{\rm r}$, {\it (c)} $\ell=\ell^{\rm cr}_{\rm r}+
    \frac{1}{16}(\ell^{\rm cr}_{\rm e}-\ell^{\rm cr}_{\rm r})$, {\it
      (d)} $\ell=\ell^{\rm cr}_{\rm e}$.  For $\ell>\ell^{\rm cr}_{\rm
      e}$, the effective potential becomes increasingly similar to
    that of the unwrapped case in figure~\ref{Veffa}{\it (b)}.  The
    effective potential is singular (exhibiting infinite repulson) at
    the singularity radius $r_{\rm r}$.  A test particle moving on the
    equatorial plane ($\theta=\pi/2$) has potential {\it (a)} for all
    values of $\ell$.  }
  \label{oblateD6D2_Veff}
\end{figure}

On the equatorial plane this is the result of ref.\cite{jmpr}, since
the solution (\ref{D6D2metric-oldcoordinates}) at
$\theta={\pi}/{2}$ reduces to the spherically symmetric solution
of $\ell=0$. The test particle feels an attractive force as it
approaches from infinity. Attraction would turn into repulsion at the
enhan\c{c}on radius, but at this point we will replace the geometry
with a physical one.

For motion along the symmetry axis, the situation is more interesting,
as it qualitatively depends on the value of the parameter $\ell$ (see
figure~\ref{oblateD6D2_Veff}). In the $\ell=0$ case the physics is
much like that described above for the equatorial motion: the
potential becomes repulsive inside $r_{\rm e}$ and this repulsion is
infinite at $r_{\rm r}$. 

For $\ell$ small there are now two enhan\c{c}on loci.
The particle can come from infinity and reach the repulsive region
just inside the outer enhan\c{c}on. This is region II.  Alternatively,
it can start from the origin (\ie, in region III, where the potential
has the same value as at infinity), and be repelled towards the inner
enhan\c{c}on shell. This is the behaviour that we noticed before
wrapping the distribution. It is {\it not} the repulson geometry which
results from wrapping.

For a test particle moving from the origin outwards, the repulson
region (resulting from wrapping) starts at the inner enhan\c{c}on.
Now, the repulsive force is directed towards the origin.  This
situation persists until $\ell=\ell_{\rm r}^{\rm cr}$.

When $\ell>\ell_{\rm r}^{\rm cr}$, a test particle moving along the
symmetry axis will still feel a repulsive force, but not an infinite
one. In principle, if it had enough energy, it could overcome the
potential barrier between inner and outer enhan\c{c}ons and move from
infinity to the origin, or {\it vice versa}. Still, because the
tension of probe branes is still negative in this region, we consider
it unphysical and will excise it.

At $\ell=\ell_{\rm e}^{\rm cr}$ the inner and outer enhan\c{c}on loci
meet and the potential barrier vanishes. However, there is still a
minimum of the potential, located at $r=\ell=r_{\rm e}$. This minimum
persists for $\ell>\ell_{\rm e}^{\rm cr}$, and is located at $r=\ell$,
as in the unwrapped case. We stress again that this remaining
repulsion has nothing to do with the wrapping, as it is the behaviour
observed in the previous section for the unwrapped geometry.


\section{Excision}
\label{excision}

Ultimately, we must remove the parts of the geometry resulting from
the wrapping which are unphysical. In order to do this,
we must see what sorts of geometry we can replace the bad parts with,
ascertaining whether it is consistent to do so. Consistency here will
be measured at the level of supergravity, bolstered by intuition from
the physics of the underlying string theory.

New stringy physics appears when the volume of the K3 gets to
$V_{\star}$.  A wrapped D6--brane probe becomes massless there and
also delocalises.  One cannot place wrapped D6--brane sources in the
regions where the volume is less than $V_{\star}$ and so, as in
previous cases, the geometry must be, to a good approximation, simply
flat space. The junction between the flat space and the well--behaved
geometry, (across which the extrinsic curvature will jump, providing a
stress tensor source) must be equivalent to a massless brane.

This is the logic that was tested in ref.\cite{jmpr} for the
spherically symmetric case with the single prototype shell, and also
with additional D2--branes, and further in ref.\cite{jm} for the case
of geometries made of mixtures of D5-- and D1--branes. In these works,
it was always spherically symmetric situation under study, but there
was a family of concentric enhan\c con shells of different types, and
sometimes non--trivial geometry was grafted in, for consistency. The
novelties here are that we have no spherical symmetry, and the nested
shells can intersect for some ranges of the parameters, making a
toroidal shape.  As we shall see, despite this complication, the
gravity junction technology\cite{darmisr} allows us to analytically
demonstrate that there is a variety of consistent excisions that we
can perform\footnote{We found the introductory sections and the
  examples presented in ref.\cite{useful} very useful for learning
  about the computation for non--spherically symmetric geometries,
  although we did not use their {\tt GR}junction package.  However, it
  may be a useful tool for projects involving more complex enhan\c con
  geometries.  See also \cite{useful2,useful3}.}.

\subsection{A Little  Hypersurface Technology}
One can perform the excision procedures that were carried out in
refs.\cite{jmpr,jm}, even though we are far from spherical symmetry,
if we are careful about how we set up the problem.

Let our spacetime $M$ have coordinates $x^\mu$, and a metric
$G_{\mu\nu}$. A general hypersurface $\Sigma$ within $M$ deserves its
own coordinates $\xi^A$, and so it is specified by an equation of the
form $f(x^\mu(\xi^A))=0$.  The unit vector normal to this hypersurface
is then specified as
\begin{eqnarray}
n_\mu^\pm=
\mp\sigma\frac{\partial f}{\partial x^\mu}\ ,\quad\mbox{where}\,\,\sigma=
\left|G^{\mu\nu}\frac{\partial f}
{\partial x^\mu}\frac{\partial f}{\partial x^\nu}\right|^{-1/2}
\ .
\labell{normal}
\end{eqnarray}
The induced metric on $\Sigma$ is the familiar
$$
G_{AB}=G_{\mu\nu}\frac{\partial x^\mu}{\partial \xi^A}
\frac{\partial x^\nu}{\partial \xi^B}\ .
$$
The main object we shall need is the extrinsic curvature of the surface. 
This is given by the pullback of the covariant derivative of the
normal vector:
\begin{eqnarray}
K_{AB}&=&\frac{\partial x^\mu}{\partial \xi^A} \frac{\partial
    x^\nu}{\partial \xi^B} \nabla_\mu n_\nu 
= -n_\mu\left(
\frac{\partial^2x^\mu}{\partial \xi^A \partial \xi^B} +
\Gamma^\mu_{\nu\rho}
\frac{\partial x^\nu}{\partial \xi^A} \frac{\partial
    x^\rho}{\partial \xi^B} \right)\ .
\labell{extrinsicmain}
\end{eqnarray}
and is a tensor in the spacetime $\Sigma$.  This might seem to be a
daunting expression, but (like many things) it simplifies a lot in
simple symmetric cases. So in the spherical case, the equation
specifying $\Sigma$ is just $f=r-R=0$, for some constant $R$,  and we can use the remaining coordinates of $M$ as coordinates on $\Sigma$. Then, $\partial
f/\partial r=1$, giving:
 $$
 n_\mu^\pm=\mp
 \frac{1}{\sqrt{\left|G^{rr}\right|}} \delta^r_\mu\ .
 $$
 using the coordinates $\xi^A=x^A$, we get the simple more commonly
 used expression:
\begin{equation}
K_{AB}^\pm=\mp {1\over2}\frac{1}{\sqrt{\left|G^{rr}\right|}}{\partial G_{AB}\over \partial r}\ .
\labell{extrinsic}
\end{equation}
In the axially symmetric case, the equation specifying $\Sigma$ is $f
= r - R(\theta) = 0$, where $R$ is now a function of $\theta$.  Since
\begin{equation}
\sigma = |\, G^{rr} + G^{\theta\theta} (\partial_\theta
R)^2|^{-1/2}\ ,
  \labell{siggy}
\end{equation}
the unit normal vectors are
\begin{equation}
n_\gamma^\pm = \mp \frac{\delta_\gamma^r - \delta_\gamma^\theta \partial_\theta R}{\sqrt{ |G^{rr} + G^{\theta\theta} (\partial_\theta R)^2|}}\ .
\end{equation}
We can compute the extrinsic curvature:
\begin{eqnarray}
K^{\pm}_{\alpha\beta} & = & \frac{1}{2}\left(n^{r \pm} \partial_{r} + n^{\theta \pm} \partial_{\theta} \right)G_{\alpha\beta}\ , \nonumber \\
K^{\pm}_{\theta\theta} & = & - \,\, n^{\pm}_r \partial_{\theta}^2 R + \frac{1}{2}( n^{r \pm} \partial_r G_{\theta\theta} - n^{\theta \pm} \partial_{\theta} G_{\theta\theta})  
 + \,\, \partial_\theta R ( - n^{r \pm} \partial_\theta G_{rr} - n^{\theta \pm} \partial_{r} G_{\theta\theta}) \nonumber \\
& & \hskip3cm + \,\,\frac{(\partial_\theta R)^2}{2} ( - n^{r \pm} \partial_r G_{rr} + n^{\theta \pm} \partial_{\theta} G_{rr})\ ,
\labell{curvature}
\end{eqnarray}
where $\alpha, \beta = t, x^{1},...,x^{6},\phi$.  This relation will
be useful below when we calculate the stress--energy tensor along
axially symmetric enhan\c con shells.

\subsection{Torodial Enhan\c con}

We will discuss the toroidal enhan\c con first.  As observed before,
the volume of K3 drops below~$V_{\star}$ in the interior of the torus.
In this region, the metric has the same form as given in
equation~\reef{D6metric-oldcoordinates}, but we replace $f_2$ and
$f_6$ with new harmonic functions, $h_2$ and $h_6$.  The precise form
of these harmonic functions will be determined by the consistency of
the theory.  Transforming to Einstein frame $G_{\mu\nu} =
e^{-{\Phi}/{2}}g_{\mu\nu}$, (this is the natural frame in which to
perform this sort of computation) we use the axially symmetric
extrinsic curvature~\reef{curvature} to determine the stress--energy
tensor along the surface joining our two solutions.

There is a discontinuity in the extrinsic curvature across the junction 
defined by
$$
\gamma_{AB} = K_{AB}^+ + K_{AB}^-\ ,
$$
The stress--energy tensor supported at this junction is given by
\cite{lanczos}:
$$S_{AB} = \frac{1}{\kappa^2}(\gamma_{AB} - G_{AB} \,\gamma_C^C)\ .$$
For our particular geometry, the stress--energy tensor can be computed
to be:
\begin{eqnarray}
S_{\mu\nu} & = &  \frac{\sigma}{2 \kappa^2} (1 + \partial_{\theta}R )\left( \frac{f_2^\prime}{f_2} + \frac{f_6^\prime}{f_6} - \frac{h_2^\prime}{h_2} - \frac{h_6^\prime}{h_6} \right)\,\, G_{\mu\nu}\ , \nonumber \\
S_{ab} & = &  \frac{\sigma}{2 \kappa^2} (1 + \partial_{\theta}R )\left(\frac{f_6^\prime}{f_6} - \frac{h_6^\prime}{h_6}\right)\,\, G_{ab}\ , \nonumber \\
S_{ij} & = &   0\ ,
\labell{SE}
\end{eqnarray}
where $\sigma$ has been defined in equation~\reef{siggy} and the prime
denotes $\partial/\partial r$. Also, Newton's constant is set by $2
\kappa^2 = 16 \pi G_N = (2\pi)^7(\alpha^{\prime})^4 g_s^2$, in the
standard units\cite{primer}. The indices $a$ and $b$ run over the K3
directions ($x^3,x^4,x^5,x^6$), while parallel to the brane's
unwrapped world--volume directions we have indices $\mu, \nu$ which
run over the ($t, x^1, x^2$). The transverse directions are labelled
by indices $(i,j)$.

As expected, the stress--energy tensor along the transverse directions
vanishes, which is consistent with the BPS nature of the system's
constituents.  The tension of the discontinuity can be obtained from
the components in the longitudinal directions.  Recall from
equation~(\ref{repulsive-potential-minimal}) that $(f_2 f_6)^\prime=0$
at the enhan\c con loci given in equation~\reef{Enhanconradius}.  For
vanishing tension, then, looking at our result in the middle line of
equation~\reef{SE}, we require that $(h_2 h_6)^\prime$ vanishes at
this radius as well.  In addition to this constraint, positive tension
between the two enhan\c con shells requires, $V(r,\theta) = V h_2/h_6
\geq V_{\star}$.  If we wish to saturate the bound and assume $h_2$
and $h_6$ have a similar form to $f_2$ and $f_6$, the harmonic
functions are in fact constant, and a suitable solution is:
\begin{eqnarray}
h_2 &=& 1 - \frac{(V - V_{\star})}{2 V}\ , \nonumber \\
h_6 &=& 1 + \frac{(V - V_{\star})}{2 V_{\star}}\ .
\end{eqnarray}
So we are able to successfully perform the task of cutting out the bad
region contained within the toroidal enhan\c con, replacing it with
flat space. It appears that we can have the resulting shell made of
zero tension branes, as is in keeping with the intuition about the
stringy fate of the constituent branes.

We must note that the ``point'' $r=0$ is not entirely satisfactory.
Indeed, we are justified in thinking of the entire geometry as that of
a torus, since in the extended coordinates, this point is really a
disc of radius $\ell$.  On the edge of this disc, as stated before,
the asymptotic volume is ambiguous, but the value $V$ seems to be the
most physically consistent, as this is what it is in the disc's
interior. There is no singularity on the disc's interior, and the
volume is not at a special value. This means that there is no
requirement to place physical branes there, and so the torus genuinely
has a hole in the centre, and not just a single point.

We have ignored the fact that the form of the harmonic function in
that region indicates that we might have a negative density of branes
on the interior of the disc. We are free to ignore this, since there
is no singularity there: the supergravity analysis tells us that we
are free to place the zero tension distribution of branes over the
whole toroidal surface instead.  Of course, there is still the fact
that there is repulsion along the symmetry axis as seen by a test particle
moving in the geometry. Again, we stress that this behaviour has
nothing to do with the wrapping: it is an artifact of the continuous
brane distribution we started with. This would not occur for other
branes, as we show in section~\ref{others}.

\subsection{Double and Oblate Enhan\c cons}
We also have the case $\ell<\ell^{\rm cr}_{\rm e}$, when the
enhan\c con lies in two disconnected parts.  We have two choices.
\begin{itemize}
\item The first is to simply cut out the entire interior region, and
  replace it by flat space. This then gives a simple oblate enhan\c
  con shape. This is perfectly fine as a solution, and has the extra
  feature that it gives a case where wrapping removes the entire disc
  located at $r=0$. This means that this is a case where the enhan\c
  con mechanism removes negative tension branes which show up in the
  unwrapped distribution.  In the event that the peculiarities
  associated to the ring at $r=0$ turn out to be unpalatable, this is
  a satisfactory conservative choice. It is also the gentlest
  generalisation from the point of view of the dual moduli space of
  the gauge theory we shall discuss in section~\ref{gaugetheory}.

\item We can also keep both the inner and the outer shell, and place
  flat space in between.  This means that the ring is kept again, but
  then the same words that we used in the case of the toroidal cases
  apply\footnote{There is a tempting possibility first raised by
    R.~C~Myers, on which we elaborate here: Perhaps the interior
    region is an inverted copy of the exterior region. This would fit
    with the observation that at $r=0$ the volume returns to the
    asymptotic value, $V$ again. The inner enhan\c con would then be
    another copy of the outer one. It would also fit with the
    observation that there is a finite repulsion: it is in fact an
    attraction in this picture. While intriguing, the fact that we can
    connect to the $\ell>\ell_{\rm cr}^{\rm e}$ case may make this
    harder to justify.  On the other hand, since $\ell$ is a parameter
    and not a physical quantity, it may be that we are free to explore
    this alternative.}.
\end{itemize}
Let us check the supergravity analysis for this case.  Recall the
three regions defined in section~\ref{wrapping}.  Region II
represented the unphysical geometry where the tension of a brane probe
becomes negative. It contains the repulson singularity.  We can
perform an incision as was done for the toroidal enhan\c con by
defining $h_2$ and $h_6$ as above, between the two enhan\c con shells.
The geometry of regions I and III are consistent and can be defined as
in equations~\reef{D6metric-oldcoordinates}.  We can alternatively
define the harmonic functions in region III to be
\begin{eqnarray}
f_2 & \to & \tilde{f}_2 = 1 - \frac{r_6^{\prime}}{r \Delta} \frac{V_{\star}}{V}\ , \nonumber \\
f_6 & \to & \tilde{f}_6 = 1 + \frac{r_6^{\prime}}{r \Delta} \ .
\end{eqnarray}
where $r_6^{\prime} \leq r_6$.  The double enhan\c con, then,
allows for some number $N^{\prime} \leq N$ of wrapped D6--branes
in the interior of our geometry.

If we perform the excision, the stress--energy tensor for the outer
shell is as in equation~(\ref{SE}) where $R$ is the outer enhan\c con
radius.  The inner shell has a stress energy tensor given by:
\begin{eqnarray}
S_{\mu\nu} & = & \frac{\sigma}{2 \kappa^2} (1 + \partial_{\theta}\tilde{R} ) \left(  \frac{h_2^\prime}{h_2} + \frac{h_6^\prime}{h_6} - \frac{\tilde{f}_2^\prime}{\tilde{f}_2} - \frac{\tilde{f}_6^\prime}{\tilde{f}_6} \right)\,\, G_{\mu\nu}\ , \nonumber \\
S_{ab} & = &  \frac{\sigma}{2 \kappa^2} (1 + \partial_{\theta}\tilde{R} ) \left(\frac{h_6^\prime}{h_6} - \frac{\tilde{f}_6^\prime}{\tilde{f}_6} \right)\,\, G_{ab}\ , \nonumber \\
S_{ij} &=& 0\ .
\end{eqnarray}
Here $\tilde{R}(\theta)$ is the equation~\reef{Enhanconradius} of the
inner enhan\c con shell defined in terms of $r_6^{\prime}$.  Again,
the transverse stresses vanish and the tension vanishes at the inner
enhan\c con shell.

So, we see that we have constructed a geometry which is consistent
with supergravity and excises the unphysical negative tension brane
geometry resulting from wrapping.

In the case of the double shell enhan\c con geometry, one might wonder
how such a geometry can be constructed.  Consider probing our geometry
with $n_2$ D2--branes as was done in refs.\cite{jm,jmpr,j}.  Such
branes are able to pass through the enhan\c con shells since they are
not wrapped on the K3 and so their tension remains positive.  This is
evident from the effective Lagrangian
\begin{equation}
{\cal L} = n_2 \tau_2 f_6 \, \, \frac{1}{2} v^2 \ .
\end{equation}
We can additionally probe our geometry with $n$ D2--D6 bound states,
as discussed in ref.\cite{jm}.  The effective Lagrangian is given by:
\begin{equation}
{\cal L} = n \tau_6 V f_2   \, \, \frac{1}{2} v^2 \ .
\end{equation}
The presence of the bound D2--brane ensures that the volume of K3 does
not drop below $V_{\star}$.  The bound state can therefore pass through the
outer enhan\c con shell and continue on to the origin.

Once we are in the interior, region III, we can separate the
D2--branes from the D6--branes since the K3 volume is greater than
$V_{\star}$ there. The D6--branes have positive tension and hence are
free to move about in the interior.  If we attempt to move a D6--brane
out of region III, the volume of K3 approaches $V_{\star}$ as the
brane nears the inner enhan\c con shell.  The D6--brane becomes
tensionless at the inner enhan\c con and melts into the shell, as
happens for brane approaching the outer shell from outside.  Hence,
D2--D6 bound states can be used to move D6--branes into the interior
and to add D6 branes to the inner enhan\c con shell.  It is easy to
imagine building the double shell enhan\c con geometry in this
fashion, starting with widely separated D2-- and D6--branes, and then
removing the D2--branes after the construction.

\section{Wrapping Other Distributed D--Branes}
\label{others}
We focused on D6--branes almost exclusively, but it should be clear
that many of the amusing features that we have seen are also present
for wrapping D4-- and D5-- branes.  In fact, we ought to stress again
here that the possibly unpalatable feature of the negative brane
density and its repulsive potential are not generic. (We have already
seen that the disc brane distributions for all of the other branes
have positive densities everywhere.)

The feature of having multiple solutions for the enhan\c con locus
will occur also. In fact, it will give up to {\it three} solutions for
the D4--brane's enhan\c con loci, but only one for the D5--brane. This
is because the harmonic functions will be of the form:
\begin{displaymath}
  f=1+\frac{r_p^{7-p}}{r^{7-p}\left(1+({\ell^2}/{r^2})\cos^2\theta\right)}\ .
\end{displaymath}
for $p=4,5$ and one non--zero parameter $\ell_1 = \ell$.  The equation
determining the enhan\c con loci is of the form of equating a ratio of
two of these $f$'s to a constant, $V_{\star}/V$ which gives a
quadratic equation in the case of the D6--brane, as we have seen, but
only a linear equation for D5--branes. The D4--brane case will give a
cubic equation.

\subsection{Unwrapped D5--Brane Distributions}

Actually, it is worth looking briefly a bit more at the D5--brane
case, keeping non--zero both of the parameters, $\ell_1, \ell_2$, that
we can have with four transverse directions. Taking the extremal
limit of the rotating black D5--brane solution given in
ref.\cite{Harmark:1999xt} we get for our unwrapped solution:
\begin{eqnarray} \label{D5metric-oldcoordinates}
ds^2 &=& f_5^{-1/2} \Big( -dt^2 + \sum_{i=1}^{5}dx_i^2 \Big)\nonumber\\
&&\hskip3cm +
f_5^{1/2} \Big( \frac{\Delta_{12}}{\Xi_1 \Xi_2} dr^2 + \Delta_{12} r^2 d \theta^2
+ \Xi_1 r^2 \sin^2\phi_1 + \Xi_1 r^2 \cos^2\phi_2 \Big)\ ,\nonumber \\
e^{\Phi} &=& f_5^{-1/2}\ ,\nonumber \\
C_6 &=& f_5^{-1} dt \wedge dx_1 \wedge \ldots \wedge dx_5\ ,
\end{eqnarray}
where
\begin{displaymath}
f_5 = 1 + \frac{r_5^2}{r^2 \Delta_{12}}\ ,
\end{displaymath}
and
\begin{displaymath}
\Delta_{12} = 1 + \frac{\ell_1^2}{r^2}\cos^2\theta + \frac{\ell_2^2}{r^2}\sin^2\theta\ ,
\qquad \Xi_1 = 1 + \frac{\ell_1^2}{r^2}\ , \qquad \Xi_2 = 1 + \frac{\ell_2^2}{r^2}\ .
\end{displaymath}
The extended coordinates are given by:
\begin{eqnarray}
y_1 & = & \sqrt{r^2 + \ell_1^2} \sin \theta \cos \phi_1\ ,\qquad
y_2  =  \sqrt{r^2 + \ell_1^2} \sin \theta \sin \phi_1\ , \nonumber \\
y_3 & = & \sqrt{r^2 + \ell_2^2} \cos \theta \cos \phi_2\ , \qquad 
y_4  =  \sqrt{r^2 + \ell_2^2} \cos \theta \sin \phi_2\ . \nonumber 
\end{eqnarray}
Considering the case $\ell_1=\ell, \ell_2=0$ will give the ring
density mentioned previously.

\begin{figure}[ht]
  \centering
  \scalebox{0.7}{\includegraphics{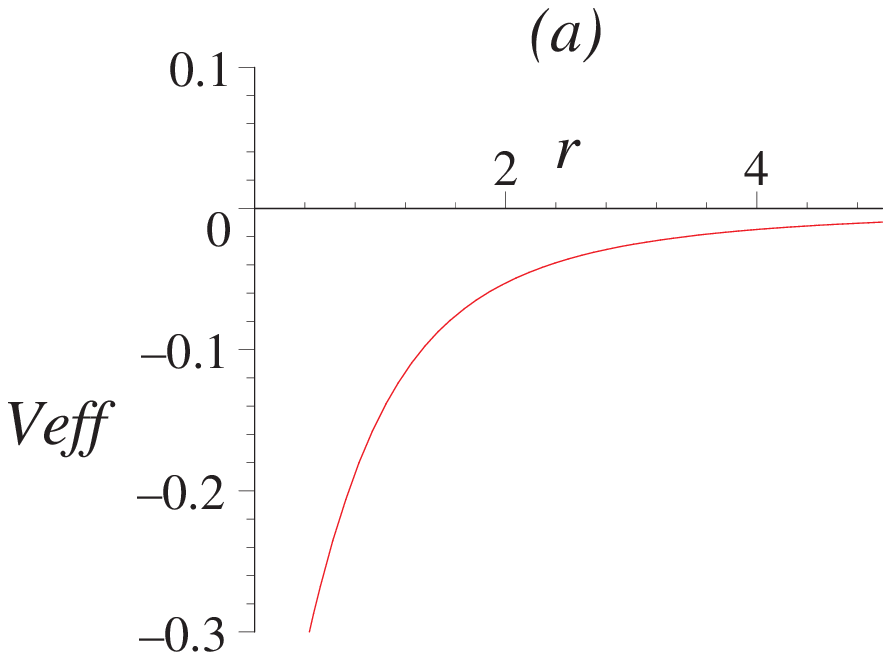}}
  \hskip1cm\scalebox{0.7}{\includegraphics{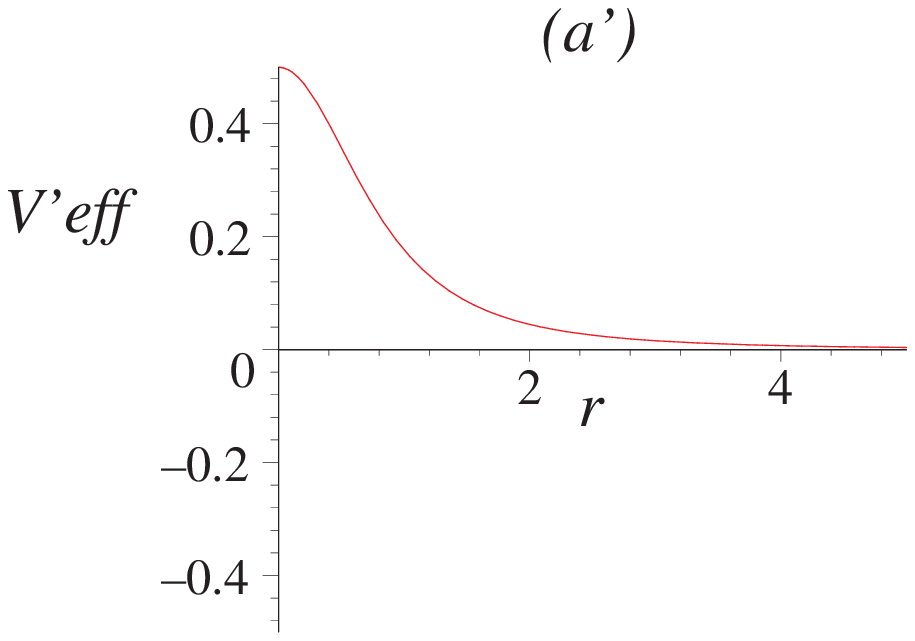}}
   \\ \vspace{2mm}
  \scalebox{0.7}{\includegraphics{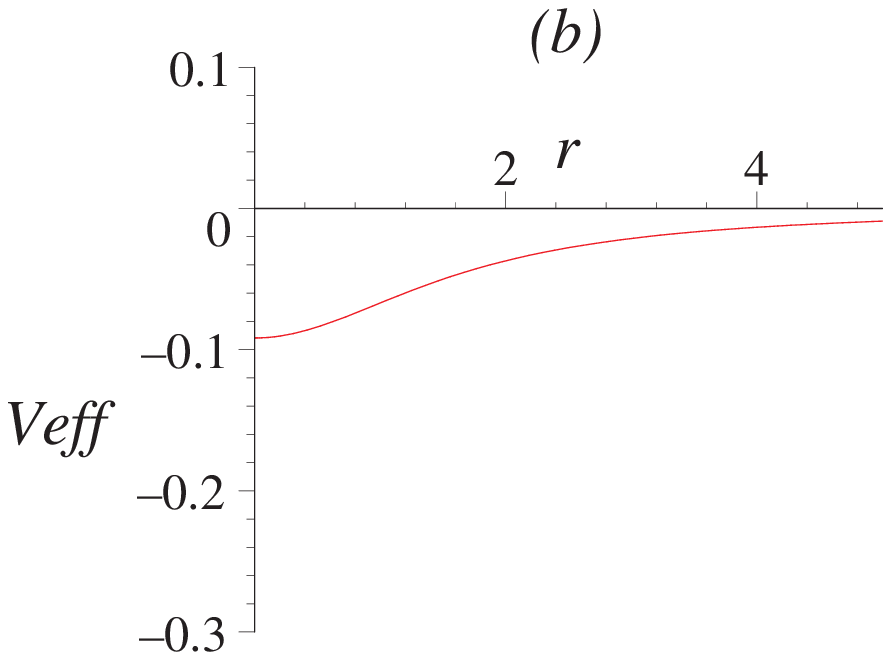}} 
  \hskip1cm\scalebox{0.7}{\includegraphics{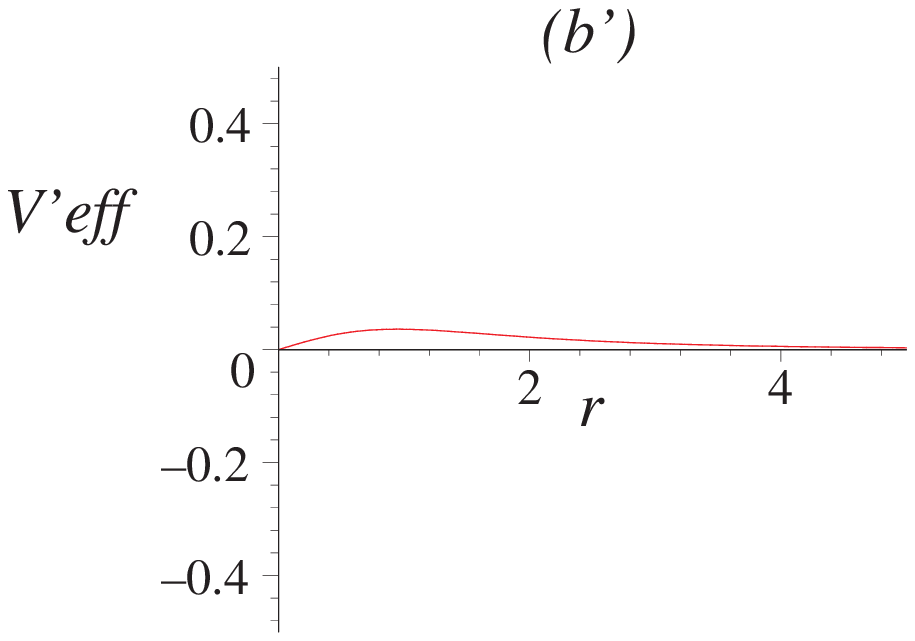}}
  \caption{\footnotesize 
    Gravitational features of the distributed D5--brane solution as
    seen by a neutral test particle.  Test particle effective
    potential (left) and its derivative (right) along the $\theta=0$
    ($\theta = \pi/2$) direction corresponding to {\it (a)} $\ell_1 = 0$
    ($\ell_2 = 0$), {\it (b)} $\ell_1 > 0$ ($\ell_2 > 0$).  The potential is
    always attractive for a particle moving in from the infinity.  }
  \label{oblateD5_Veff}
\end{figure}

Let us consider particle probes of the geometry.  Previously we
associated repulsive gravitational features of the D6--brane
distribution with a negative brane density on the disc.  The D5--brane
distribution did not have negative brane density, and indeed, does not
show these repulsive features. In this case the transverse space--time
has one more dimension, but it is also endowed with one additional
Killing vector related to the $\phi_2$ coordinate.  Therefore an
analysis similar to the one set forth in section~\ref{probing} shows
that test particle motion in $\theta = 0$ and $\theta = \pi/2$
directions is governed by the same effective potential given in
equation~(\ref{veef}). In fact, the effective potential in the $\theta
= 0$ direction does depend only on $\ell_1$ and in the $\theta =
\pi/2$ direction only on $\ell_2$, while the form of dependence is
identical.  As figure~\ref{oblateD5_Veff} shows, the repulsive
gravitational features are completely absent.

\subsection{Wrapped D5--Brane Distributions}

\begin{figure}[ht]
  \centering
  \includegraphics{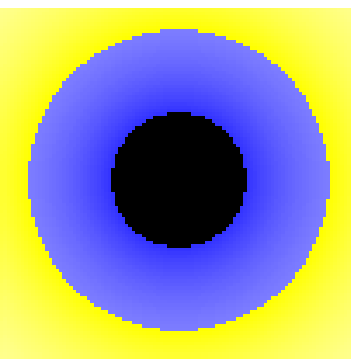}
  \includegraphics{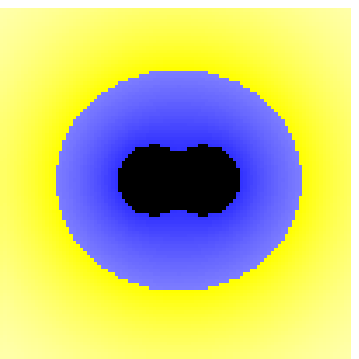}
  \includegraphics{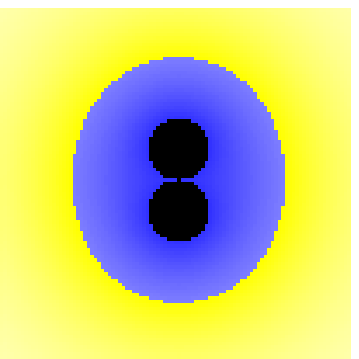}
  \includegraphics{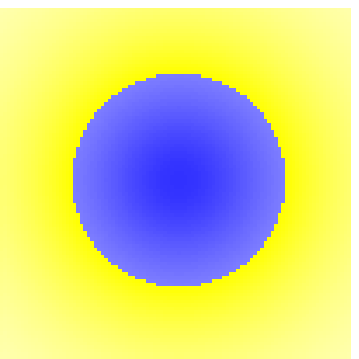}
  \\ \vspace{2mm}
  \includegraphics{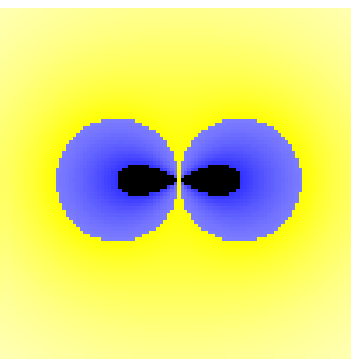}
  \includegraphics{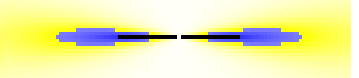}
  \includegraphics{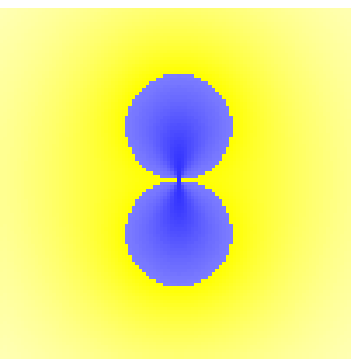}
  \includegraphics{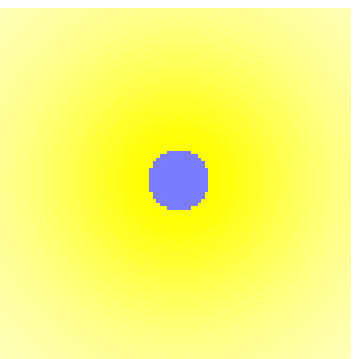}
  \caption{\footnotesize 
    Some two--dimensional slices through the non--spherical wrapped
    D5--brane geometry before excision.  $\theta=0$ corresponds to the
    vertical direction and $\theta=\pm {\pi}/{2}$ to the horizontal.
    The configuration is symmetric in $\phi_1$ and $\phi_2$ (not shown
    on the figure).  The colour coding is as in
    figure~\ref{K3volume-oldcoordinates}. The shapes vary for
    different values of the parameters $\ell_1$ and $\ell_2$: $\ell_1
    = \ell_2 = 0$ (top, left), $\ell_1 = \frac{7}{8}\ell^{\rm cr}_{\rm
      r}$, $\ell_2 = 0$ (top, second), $\ell_1 = 0$, $\ell_2 =
    \ell^{\rm cr}_{\rm r}$ (top, third), $\ell_1 = \ell_2 = \ell^{\rm
      cr}_{\rm r}$ (top, right), $\ell_1 = \ell^{\rm cr}_{\rm e}$,
    $\ell_2 = 0$ (bottom, left) $\ell_1 = 8 \ell^{\rm cr}_{\rm e}$,
    $\ell_2 = 0$ (bottom, second) $\ell_1 = \ell^{\rm cr}_{\rm r}$,
    $\ell_2 = \ell^{\rm cr}_{\rm e}$ (bottom, third) $\ell_1 = \ell_2
    = \ell^{\rm cr}_{\rm e} - \frac{1}{16}(\ell^{\rm cr}_{\rm
      e}-\ell^{\rm cr}_{\rm r})$ (bottom, right).  }
  \label{oblateD5D1_K3volume}
\end{figure}

Wrapping the  D5--brane configuration (\ref{D5metric-oldcoordinates}) on
K3 yields
\begin{eqnarray}
ds^2 &=& f_5^{-1/2} f_1^{-1/2} \big( -dt^2 + dx^2 \big)
+ f_5^{-1/2} f_1^{1/2} V ds_{K3}^2 \nonumber \\
&& + f_5^{1/2} f_1^{1/2} \Big( \frac{\Delta_{12}}{\Xi_1 \Xi_2} dr^2 + \Delta_{12} r^2 d \theta^2
+ \Xi_1 r^2 \sin^2\phi_1 + \Xi_1 r^2 \cos^2\phi_2 \Big)\ , \nonumber \\
e^{\Phi} &=& f_5^{-1/2} f_1^{1/2}\ ,\nonumber  \\
C_6 &=& f_5^{-1} dt \wedge dx \wedge V ds_{K3}\ , \nonumber \\
C_2 &=& f_1^{-1} dt\ , \wedge dx
\end{eqnarray}
where
\begin{displaymath}
f_5 = 1 + \frac{r_5^2}{r^2 \Delta_{12}}\ , \qquad
f_1 = 1 - \frac{r_1^2}{r^2 \Delta_{12}}\ , \qquad
r_1^2 = \frac{V_{\star}}{V}r_5^2\ ,
\end{displaymath}
and $\Delta_{12}$, $\Xi_1$, $\Xi_2$, \etc., defined as before.
There is a repulson singularity at
\begin{displaymath}
r_{\rm r} = \sqrt{\frac{r_5^2 V_{\star}}{V} \Big( 1 -
\frac{V}{r_5^2 V_{\star}} (\ell_1^2 \cos^2\theta + \ell_2^2 \sin^2\theta) \Big) }\ ,
\end{displaymath}
where the running volume of K3 shrinks to zero. Interestingly, the
singularity disappears completely when both parameters $\ell_1$ and
$\ell_2$ exceed the critical value
\begin{displaymath}
\ell^{\rm cr}_{\rm r} = \sqrt{\frac{r_5^2 V_{\star}}{V}}\ .
\end{displaymath}
The singularity is always surrounded by the enhan\c{c}on shell at
\begin{displaymath}
r_{\rm e} = \sqrt{\frac{2 r_5^2 V_{\star}}{V - V_{\star}} \Big( 1 -
\frac{V - V_{\star}}{2 r_5^2 V_{\star}} (\ell_1^2 \cos^2\theta + \ell_2^2 
\sin^2\theta) \Big) }\ . 
\end{displaymath}
Depending on the values of $\ell_1$ and $\ell_2$, this  enhan\c{c}on assumes
various shapes, but is always connected. Again, it is interesting that  when
both parameters $\ell_1$ and $\ell_2$ exceed the critical value
\begin{displaymath}
\ell^{\rm cr}_{\rm e} = \sqrt{\frac{2 r_5^2 V_{\star}}{V - V_{\star}}}\ ,
\end{displaymath}
the enhan\c{c}on disappears completely. See
figure~\ref{oblateD5D1_K3volume} for a series of snapshots.
It is important to note that the following result is true
\begin{equation}
\frac{\partial}{\partial r} \Big( f_5 f_1 \Big) \Big|_{r=r_{\rm e}} = 0\ .
\end{equation}
It is this that will ensure that the supergravity matching computation
will go through in a similar manner as in the D6--branes case here,
and as in the D5--brane cases studied in ref.\cite{jm}.

\begin{figure}
  \centering
  \scalebox{0.7}{\includegraphics{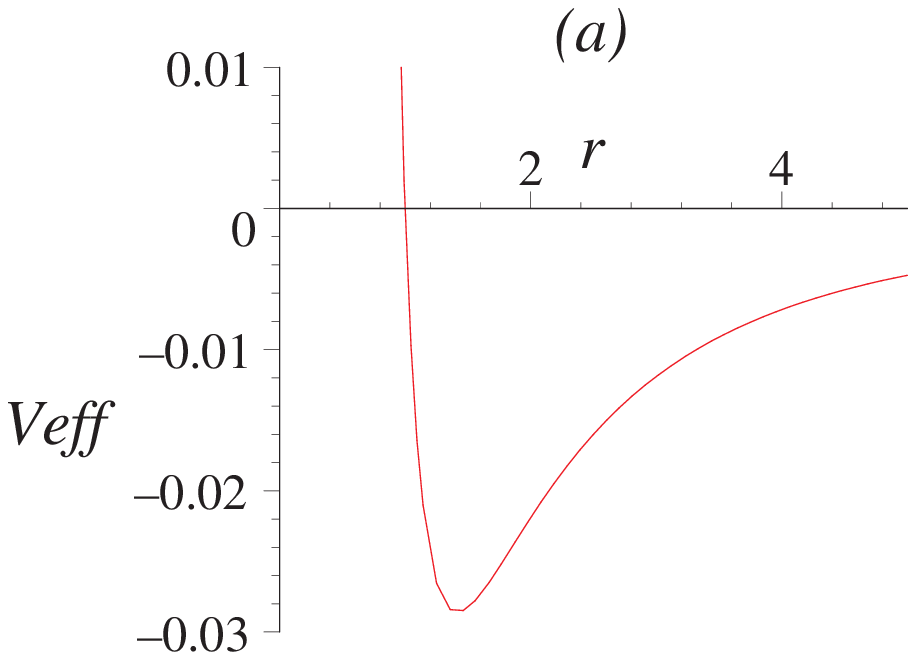}}
  \hskip1cm\scalebox{0.7}{\includegraphics{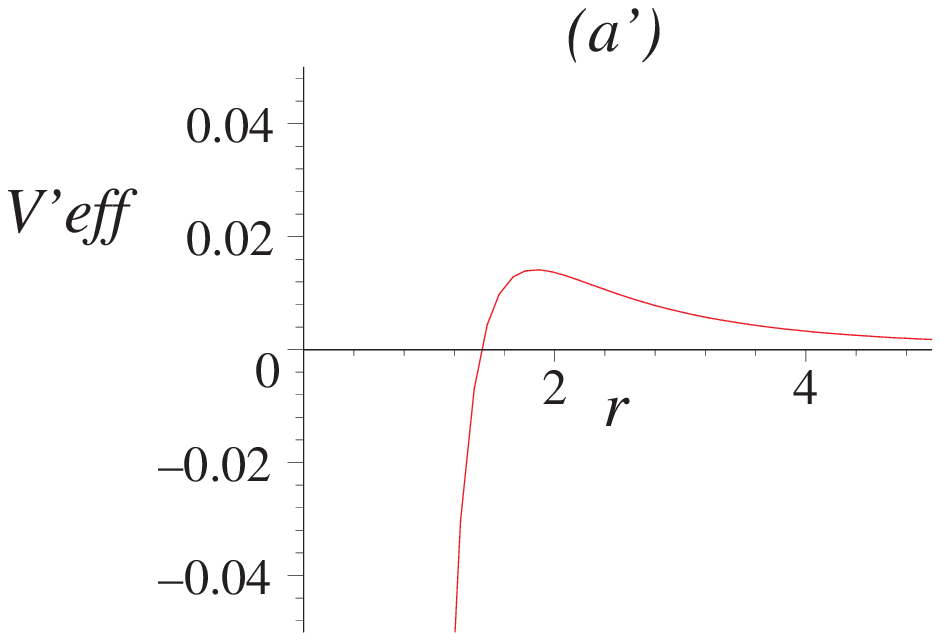}}
\scalebox{0.7}{\includegraphics{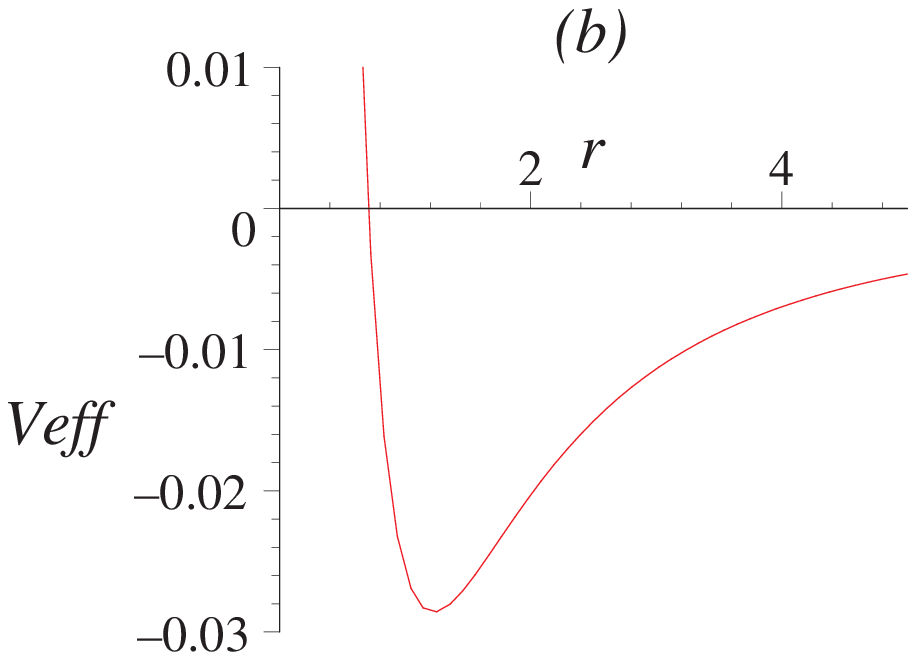}}
 \hskip1cm\scalebox{0.7}{\includegraphics{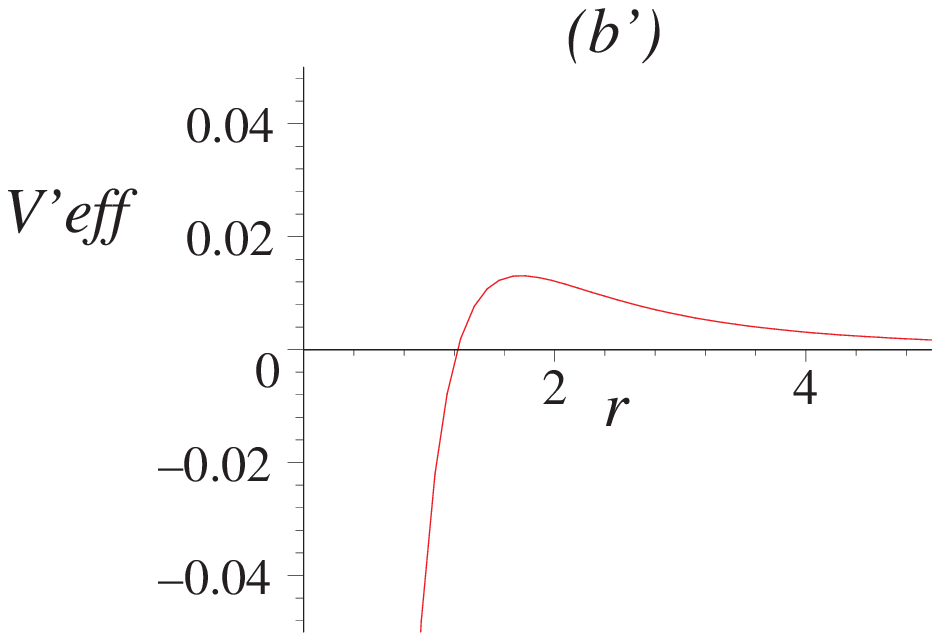}}
 \\ \vspace{2mm}
   \scalebox{0.7}{\includegraphics{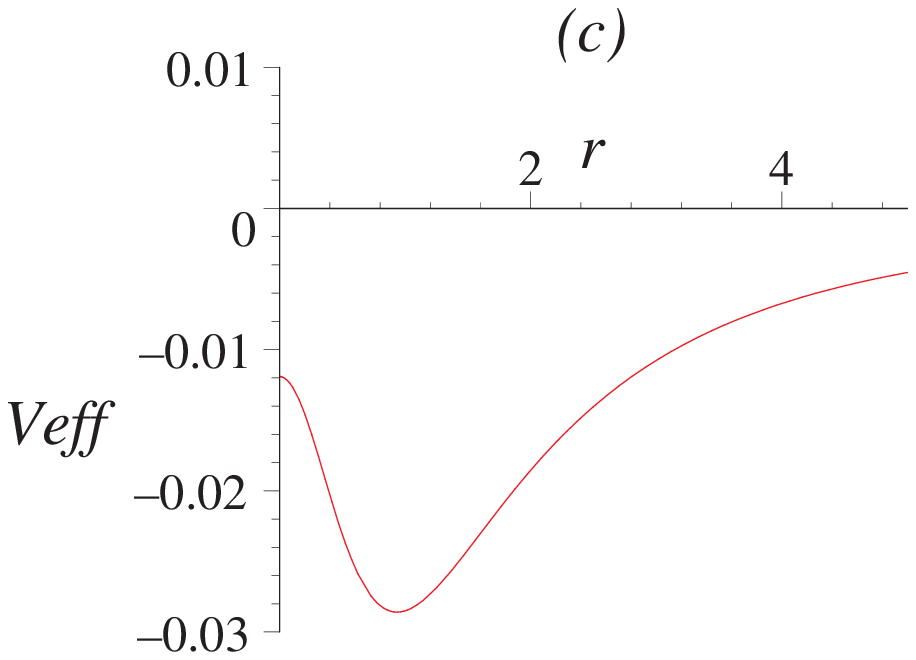}}
  \hskip1cm
  \scalebox{0.7}{\includegraphics{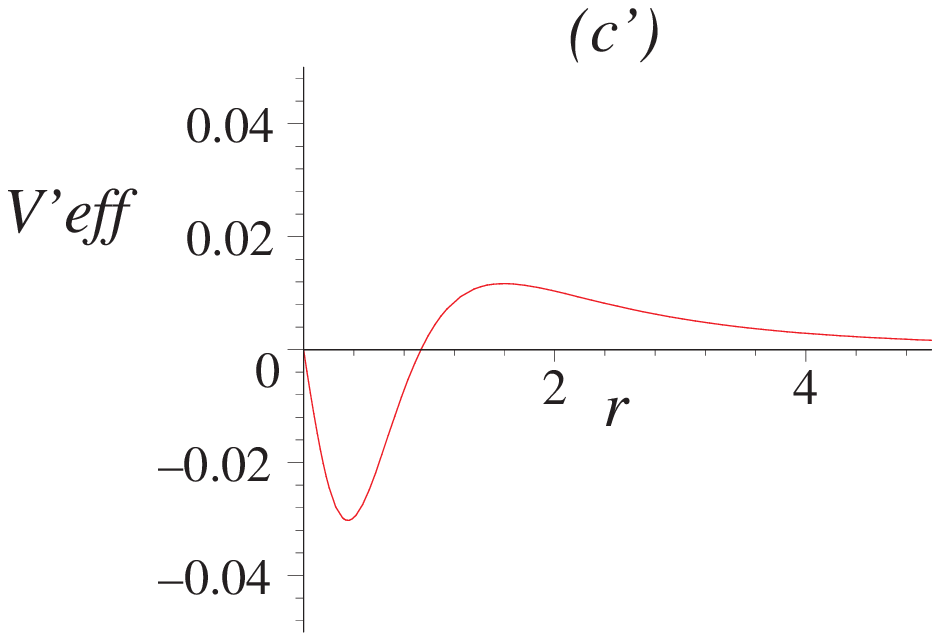}}
\scalebox{0.7}{\includegraphics{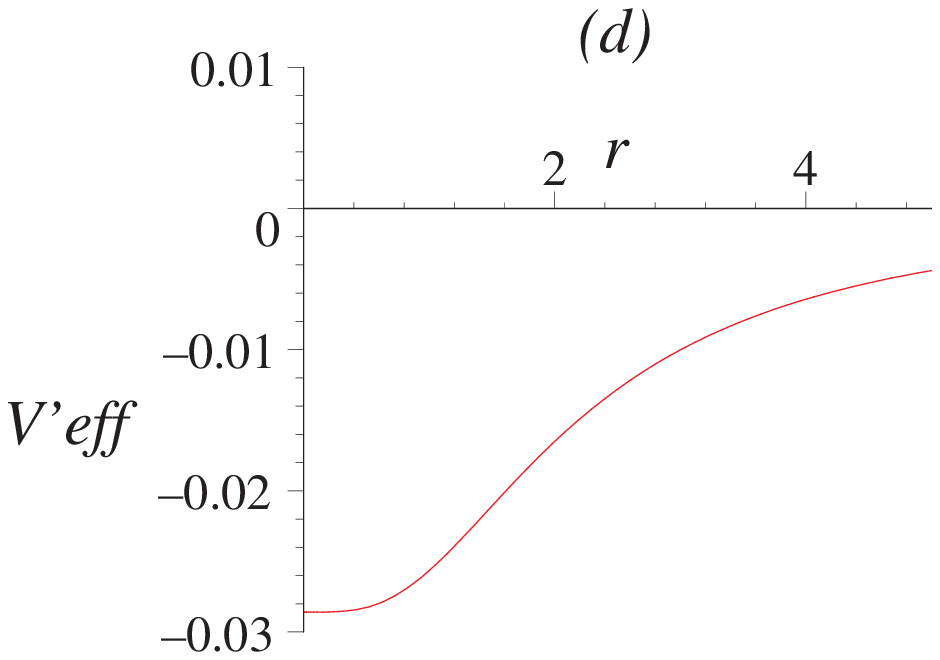}}
  \hskip1cm\scalebox{0.7}{\includegraphics{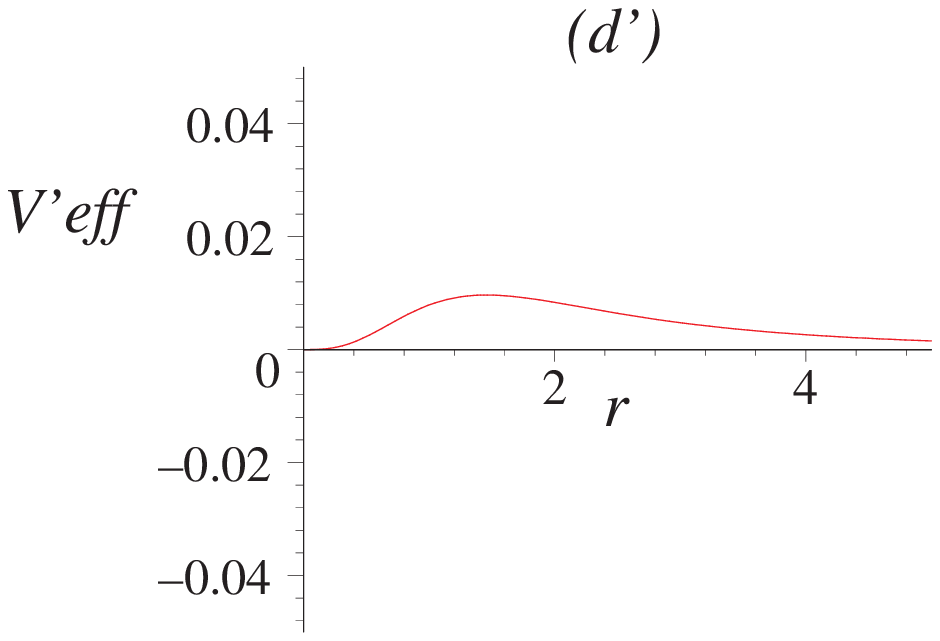}}
  \caption{\footnotesize 
    Gravitational features of the distributed D5--brane solution
    wrapped on K3 as seen by a neutral test particle.  Test particle
    effective potential (left) and its derivative (right) along the
    $\theta=0$ ($\theta = \pi/2$) direction corresponding to {\it (a)}
    $\ell_1 = 0$ ($\ell_2 = 0$), {\it (b)} $\ell_1=\ell^{\rm cr}_{\rm r}$
    ($\ell_2=\ell^{\rm cr}_{\rm r}$), {\it (c)} $\ell_1=\ell^{\rm cr}_{\rm
      r}+ \frac{1}{2}(\ell^{\rm cr}_{\rm e}-\ell^{\rm cr}_{\rm r})$,
    ($\ell_2=\ell^{\rm cr}_{\rm r}+ \frac{1}{2}(\ell^{\rm cr}_{\rm
      e}-\ell^{\rm cr}_{\rm r})$), {\it (d)} $\ell_1=\ell^{\rm cr}_{\rm e}$
    ($\ell_1=\ell^{\rm cr}_{\rm e}$).  Effective potential is singular
    (exhibiting infinite repulson) at $r_{\rm r}$ and has a minimum
    at~$r_{\rm e}$.  }
  \label{oblateD5D1_Veff}
\end{figure}

Let us consider particle probes again, in order to check where the
repulsive regions are. In contrast to the D6--brane case, we do not
expect any repulsive features to arise which are not attributable to
the wrapping. The presence of the repulson singularity is signalled by
a singularity in the effective potential occurring at $r_{\rm r}$.
This is surrounded by the enhan\c{c}on shell, residing precisely at
the minimum of the potential, {\it i.e.,} where the repulsive region
starts. (See figure~\ref{oblateD5D1_Veff}.)  There is still some
residual finite repulsion even after $\ell_1$ ($\ell_2$) exceeds
$\ell^{\rm cr}_{\rm r}$ and the repulson singularity disappears
(figure~\ref{oblateD5D1_Veff}{\it (c)}). Beyond $\ell_1>\ell^{\rm
  cr}_{\rm e}$ ($\ell_2>\ell^{\rm cr}_{\rm r}$) the effective potential
becomes completely attractive with a minimum at the origin $r=0$.

\section{Gauge Theory}
\label{gaugetheory}
It is worth remembering that some of our results pertain to gauge
theory. Wrapping the D6--branes on the K3 results in an ${\cal N}=4$
$(2+1)$--dimensional gauge theory with UV gauge
coupling\cite{jpp,primer}:
$$
g^2_{\rm YM}=(2\pi)^4g_s \alpha^{\prime3/2} V^{-1}\ .
$$
One of the results that one can readily extract from the analysis
of the spherical case\cite{jpp} is  the piece of the
gauge theory moduli space corresponding to the movement of a single
wrapped probe.  In gauge theory, this is the metric derived from the
kinetic terms in the low energy effective action for the field
breaking $SU(N)\to SU(N-1)\times U(1)$. The probe computation yields
the tree level and one--loop contribution to this result, and there
are no perturbative corrections.

In the present case, we should be able to see new physics, since the
parameter $\ell$ should have some meaning in the gauge theory. In
fact, we expect that it should be controlling the vacuum expectation
value of an operator made by the symmetric product of the adjoint
scalars, by analogy with the case in the AdS/CFT\cite{adscft,adscft2}
context.  Here, the $R$--symmetry is $SU(2)\simeq SO(3)$. The three
adjoint scalars in the gauge multiplet, transforming as the ${\bf 3}$,
can be combined by symmetric product. The $R$--charge of the resulting
operator is computed in the usual way. For example, ${\bf 3}\times{\bf
  3}$=${\bf 1}+{\bf 5}+{\bf 3}$, where the ${\bf 5}$ is the symmetric
traceless part. This is of course more familiar as the $j=2$ case in
the standard angular momentum notation where the irreducible
representations of the $R$--symmetry group are written as $2j+1$
dimensional. We should see some sign of this show up here, and indeed
we do.

\subsection{Metric on Moduli Space}
We probe the {\it repaired} geometry with a single wrapped D6--brane
in order to extract the data of interest.  After including the $U(1)$
world--volume gauge sector into the probe calculation, dualising the
gauge field to get the extra periodic scalar\cite{jpp,primer}, the
kinetic term of the effective action in the non--flat regions becomes
\begin{displaymath}
T = h(r,\theta) \,\,\left(\frac{\Delta}{\Xi} \,{\dot{r}^2} 
+ \Delta r^2 \,\dot{\theta}^2\ + \Xi \,r^2 \sin^2\theta \,\dot{\phi}^2 \right) 
+  h(r,\theta)^{-1} \left( \dot{s}/2 - \mu_2 C_{\phi} \dot{\phi}/2\,\right)^2 .
\end{displaymath}
where $h(r,\theta)$ is defined by
\begin{displaymath}
h(r,\theta) = \frac{\mu_6}{2}\,( V f_2 - V_{\star} f_6 )
\end{displaymath}
and $C_{\phi}$ is the magnetic potential corresponding to  the D6--brane
charge
\begin{displaymath}
C_{\phi} = - r_6 \ \frac{\Xi}{\Delta} \cos \, \theta\ .
\end{displaymath}
To extract the gauge theory we work with variables\cite{adscft}:
\begin{displaymath}
U = \frac{r}{\alpha^{\prime}}\ , \quad a = \frac{l}{\alpha^{\prime}}\ ,
\end{displaymath}
and take the decoupling limit, which involves holding $U,\ a,$ and
$g_{YM}^2$ fixed while taking $\alpha^{\prime} \to 0$. The metric becomes:
\begin{displaymath}
ds^2 = h(U,\theta) \ \left( \frac{\Delta^{\prime}}{\Xi^{\prime}} 
\,dU^2 + \Delta^{\prime} U^2 \, d\theta^2\ + \Xi^{\prime} \,
U^2 \sin^2\theta \,d\phi^2 \right) +  h(U,\theta)^{-1} 
\left( d\sigma - \frac{N}{8 \pi^2} A_{\phi} d\phi\,\right)^2\ ,
\end{displaymath}
with 
\begin{displaymath}
h(U,\theta) = \frac{1}{8 \pi^2 g^{2}_{YM}} \left( 1 
- \frac{g^{2}_{YM} N}{U \Delta^{\prime}}\right)\ .
\end{displaymath}
Here we have defined
\begin{displaymath}
\Delta^{\prime} = 1  +  \frac{a^2}{U^2} \cos^2 \theta \ , \quad \Xi^{\prime} = 1 + \frac{a^2}{U^2}\ ,
\end{displaymath}
and 
\begin{displaymath}
\sigma = \frac{s \alpha^{\prime}}{2}\ , \quad A_{\phi} = -\frac{1}{2} \frac{\Xi^{\prime}}{\Delta^{\prime}} \cos \theta\ .
\end{displaymath}
This might not seem terribly inspiring, but recall that we can work in
the extended coordinate system given by the obvious generalisation of
equation~\reef{extendedcoordinates}:
\begin{eqnarray} 
W_1 & = & \sqrt{U^2 + a^2} \, \sin \theta \, \cos \phi \nonumber \\
W_2 & = & \sqrt{U^2 + a^2} \, \sin \theta \, \sin \phi \nonumber \\
W_3 & = & U \, \cos \theta . \labell{extendedcoordinates2}
\end{eqnarray}
In terms of these, our moduli space result is the standard Taub--NUT form:
\begin{equation}
ds^2 = H(W,{\hat \theta}) \ \left(d\vec{W}\cdot d\vec{W}\right) 
+  H(W,{\hat \theta})^{-1} 
\left(d\sigma - \frac{N}{8 \pi^2} A_{\phi} d{\phi}\,\right)^2\ ,
\end{equation}
with 
\begin{displaymath}
H(W,{\hat \theta}) = \frac{1}{8 \pi^2 g^{2}_{YM}} \left( 1 
- \frac{g^{2}_{YM} N  \sqrt{\Lambda + \Sigma} }{\sqrt{2}\Sigma}\right)\ .
\end{displaymath}
where
\begin{displaymath}
\Sigma = \sqrt{\Lambda^2 + 4 a^2 W_3^2}, \qquad
\Lambda = W^2 - a^2, \qquad
W = \sqrt{W_1^2 + W_2^2 + W_3^2}\ .
\end{displaymath} 
The content is in the harmonic function, and we can expand it for large
$W$  using our earlier observations in equations~\reef{Legendre}:
\begin{eqnarray}
H & = & \frac{1}{8 \pi^2 g^{2}_{YM}} \left( 1 - \frac{g^{2}_{YM} N  }{W} 
\sum_{n=0}^{\infty} (-1)^n
\left(\frac{a}{W}\right)^{2n} P_{2n}(\cos\ \hat{\theta}) \right)\ ,
\labell{Operators}
\end{eqnarray}
where we have defined new polar coordinate angles in an analogous
manner to that shown in equation~\reef{newangles}, and the $P_{2n}(x)$
are the Legendre polynomials in $x$, as before (see
equation~\reef{normalised}).

The leading terms in this large $W$ expansion should have an
interpretation as the contribution of the operators which are switched
on. The $n=0$ result is that of the spherical case\cite{jpp}. The case
$n=1$ comes with the Legendre polynomial $P_2(\cos{\hat\theta})$,
which has the $R$--charge of the~${\bf 5}$, the simplest operator one
can make out the adjoint scalars. So the parameter $\ell$, (\ie, $a$)
controls the vacuum expectation value of this operator, with
subleading contributions coming from the higher spherical harmonics.

Notice that this expansion is not sensitive to some of choices that we
can make in doing the excision. In particular, while it works for any
$\ell$, it is for large $y$. So from the point of view of this
expansion, all of the choices are equivalent to the excisions which
give a single locus: For $\ell<\ell_{\rm e}^{\rm cr}$, this is the
oblate enhan\c con with flat space inside, while for $\ell>\ell_{\rm
  e}^{\rm cr}$, it is the toroidal shape.  The difference between the
two is presumably non--perturbative in the operator expansion. It
would be interesting to find gauge theory meaning for cases where we
can choose to have a double shell.

\section{Concluding Remarks}
\label{conclude}

We have seen that enhan\c con shapes quite different from the
prototype spherical case can be well described within supergravity. We
achieved this by wrapping distributions of D--branes on~K3, and
uncovered many interesting and beautiful features of the resulting
geometry.

While the main example we used here (the D6--brane distribution) had a
physical oddity at its core (a negative contribution to the brane
density), we continued to use it as out main example, since it is
clear that this feature does not affect the discussion of the enhan\c
con. Indeed, we showed that all of the salient features we wanted to
illustrate are present for D5--brane distributions, (and very likely
also extend to D4--branes) which do not have any such problems with
the distribution's density.  For the D5--branes in fact, it is clear
that the excision process always removes the original disc
distribution of branes (this time they are all on the edge), and so
the problem in that case is moot.

In fact, it would be interesting to study our D6--brane distribution
further to determine if there is some mechanism, perhaps similar to
the enhan\c con, which might resolve the apparently unphysical
features (repulsive force and negative brane density) that we
observed.  This odd behaviour may be a residual effect inherited from
the fully rotating configuration since we are studying an extremal
limit of that solution.  Perhaps a consideration of the full solution
will provide additional insight.

We were able to use the geometry to extract some quite interesting new
information about the associated $SU(N)$ gauge theory whose moduli
space is isomorphic to that of the wrapped D6--brane problem. One can
read off the details of which operator vacuum expectation values have
been switched on from an expansion of a probe result in terms of
spherical harmonics. Of course, similar results for other gauge groups
should be easily extracted by a generalisation of the work of
ref.\cite{jj} along the lines done here.

Overall, it is very satisfying that we can find and exactly analyse
such intricate structures, successfully correcting poorly behaved
supergravity geometries with knowledge from the underlying string
theory. This is encouraging, since such studies have potential
applications to searches for realistic gauge/geometry duals, better
understanding of singularities in string theory, and a host of other
inter--related problems.

\section*{Acknowledgements}
L.D. would like to thank A. Lerda for useful discussions during the
initial stages of this research. L.D. also thanks the Department of
Mathematical Sciences, University of Durham for their hospitality and
for a Grey College visiting Fellowship. We thank R. C. Myers for a
comment.  This paper is report number MIT-CTP-3225 and DCPT-01/77.

\end{document}